\documentclass[11pt]{article}
	
\newcommand{\blind}{0}

\makeatletter
\renewcommand\section{\@startsection {section}{1}{\z@}%
                                   {-3.5ex \@plus -1ex \@minus -.2ex}%
                                   {2.3ex \@plus.2ex}%
                                   {\normalfont\fontfamily{phv}\fontsize{16}{19}\bfseries}}
\renewcommand\subsection{\@startsection{subsection}{2}{\z@}%
                                     {-3.25ex\@plus -1ex \@minus -.2ex}%
                                     {1.5ex \@plus .2ex}%
                                     {\normalfont\fontfamily{phv}\fontsize{14}{17}\bfseries}}
\renewcommand\subsubsection{\@startsection{subsubsection}{3}{\z@}%
                                    {-3.25ex\@plus -1ex \@minus -.2ex}%
                                     {1.5ex \@plus .2ex}%
                                     {\normalfont\normalsize\fontfamily{phv}\fontsize{14}{17}\selectfont}}
\makeatother

\usepackage{amsmath}
\usepackage{graphicx}
\usepackage{tabularx}
\usepackage{enumerate}
\usepackage{natbib} 
\usepackage{url} 
\usepackage{xcolor}
\usepackage{amssymb}
\usepackage{threeparttable}
\usepackage{multirow}

\usepackage{graphicx, amsmath, amsfonts, xcolor, tcolorbox,soul, titlesec, hyperref,algorithm, algpseudocode, verbatim, enumitem, float, wrapfig, todonotes, setspace, caption, bm, tabularx, enumitem}
\usepackage[letterpaper, margin=1in]{geometry}
\counterwithout{figure}{section}

\setlist[itemize]{itemsep=0pt, leftmargin=*}
\setlist[enumerate]{topsep=0pt,itemsep=0ex,partopsep=1ex,parsep=1ex, leftmargin=*}
\titlespacing\section{0pt}{*1}{*0}
\titlespacing\subsection{0pt}{*1}{*0}
\titlespacing\subsubsection{0pt}{*1}{*0}
\titlespacing\paragraph{0pt}{*0.6}{*0}

\usepackage[skip=2pt plus1pt, indent=18pt]{parskip}

\begin{document}
    
    \def\spacingset#1{\renewcommand{\baselinestretch}%
        {#1}\small\normalsize} \spacingset{1}
    
    \if0\blind
    {
        \title{\bf \emph {Prognostics of Multisensor Systems with Unknown and Unlabeled Failure Modes via Bayesian Nonparametric Process Mixtures}}
        \author{Kani Fu$^a$, Sanduni S Disanayaka Mudiyanselage$^b$, Chunli Dai$^b$ and Minhee Kim$^{a*}$\\
        $^a$Department of Industrial and Systems Engineering, University of Florida, Gainesville, USA \\
         $^b$School of Forest, Fisheries, and Geomatics Sciences, University of Florida, Gainesville, USA }
        \date{}
        \maketitle
    } \fi
    
    \if1\blind
    {

        \title{\bf \emph{Prognostics of Multisensor Systems with Unknown and Unlabeled Failure Modes via Bayesian Nonparametric Process Mixtures}}
        \author{Author information is purposely removed for double-blind review}
        \date{}
        \maketitle
    } \fi
    \bigskip

\begin{abstract}

Modern manufacturing systems often experience multiple and unpredictable failure behaviors, yet most existing prognostic models assume a fixed, known set of failure modes with labeled historical data. This assumption limits the use of digital twins for predictive maintenance, especially in high-mix or adaptive production environments, where new failure modes may emerge, and the failure mode labels may be unavailable. 

To address these challenges, we propose a novel Bayesian nonparametric framework that unifies a Dirichlet process mixture module for unsupervised failure mode discovery with a neural network-based prognostic module. The key innovation lies in an iterative feedback mechanism to jointly learn two modules. These modules iteratively update one another to dynamically infer, expand, or merge failure modes as new data arrive while providing high prognostic accuracy.

Experiments on both simulation and aircraft engine datasets show that the proposed approach performs competitively with or significantly better than existing approaches. It also exhibits robust online adaptation capabilities, making it well-suited for digital-twin-based system health management in complex manufacturing environments. The code will be made publicly available on GitHub upon the publication of this paper.

\end{abstract}

\noindent%
{\it Keywords:} Failure mode identification; RUL prediction; Predictive maintenance; Digital twin.

\spacingset{1.5} 

\section{Introduction}
\label{sec:introduction}

Unexpected failures in manufacturing systems can have significant consequences, including production downtime, safety hazards, and logistical disruptions. To reduce these risks, prognostics aims to accurately predict the remaining useful lifetime (RUL) of a given system based on its current and historical condition monitoring observations \citep{liu2013data}. 
Driven by advances in sensing technologies, modern manufacturing systems are increasingly equipped with multiple sensors to capture more comprehensive information about real-time system health status \citep{kim2021adaptive}. Prognostics based on multisensor signals has been widely applied across various industries, such as aircraft engines \citep{saxena2008damage}, batteries \citep{goebel2008prognostics}, and electrical systems \citep{liu2018patent}. Despite these advances, accurate prognostics remain challenging due to increasingly complex failure mechanisms and operating conditions.

Traditional prognostic studies typically assume that systems operate under a single failure mode \citep{Li2025view, kim2019generic}. In reality, however, manufacturing systems frequently experience multiple failure modes, each originating from different physical degradation processes. For example, aircraft engines commonly experience separate failure modes related to different components, such as the high-pressure compressor and the fan \citep{saxena2008damage}. Real-world applications indicate that these various failure modes significantly influence degradation trajectories and the lifespan of systems. As a result, prognostic models that assume a single failure mode may lack sufficient accuracy in such domains. Integrating failure mode identification with system prognostics is thus essential for comprehensive and effective health management.

Motivated by these needs, recent literature has begun incorporating multi-failure-mode information into prognostics  \citep{li2022deep, wang2024, fu2025degradation}. While some existing approaches represent important advancements, they often rely on strong assumptions: (i) the number of failure modes is fixed and known in advance, and (ii) the failure modes of historical (failed) systems are available. Both assumptions are often violated in non-stationary and complex systems. In high-mix, low-volume, or rapidly evolving manufacturing environments \citep{tomavsevic2021lean,wan2020artificial}, new failure modes can emerge under unseen requirements, environments, or workloads. In this study, ``evolving (emerging)'' failure modes refer to the possibility that new failure modes may appear in the population over time, rather than an individual system switching modes during its life. In addition, systems that fail in the field are often simply replaced or discarded without thorough failure analysis due to cost, time, or operational constraints, leaving failure modes (labels) unknown. These realities create a critical gap for failure mode identification and prognostics.

To address these limitations that presuppose fixed and fully labeled failure modes, we develop a generalizable and adaptive failure mode identification and prognostics framework that is suited to multi-sensor systems operating in complex and non-stationary environments. The proposed framework jointly addresses failure mode identification and RUL prediction through an integrated modeling architecture. A Dirichlet Process Mixture Model (DPMM) performs unsupervised failure mode discovery, while a neural network-based prognostics module predicts RUL based on both diagnostic information and multi-sensor data. The key novelty lies in how these two components iteratively update and `mutually' inform each other, continuously refining and improving the prognostic accuracy. The resulting model is capable of dynamically expanding to incorporate newly emerging failure modes or merging redundant ones.

In summary, the main contributions of this study are as follows:
\begin{itemize}
    \item We propose a novel Bayesian nonparametric methodology to integrate a DPMM for unsupervised failure mode identification with a neural network-based RUL prognostics in a unified manner.
    \item Our method establishes an iterative feedback mechanism between failure mode identification and prognostics.
    \item The proposed model is capable of inferring the number of failure modes from unlabeled historical (failed) systems and updating it online under non-stationary system dynamics.
    \item Unlike existing works, our proposed method does not assume knowledge of types and number of failure modes in advance, which provides a more general and practically applicable solution.
    \item Through comprehensive simulation and real-world case studies, we demonstrate the efficiency, flexibility, and robustness of the proposed framework in comparison with existing methods.
\end{itemize}

The remainder of this paper is organized as follows:
Section~\ref{sec:literature} reviews the existing literature in system prognostics with or without failure mode identification.
Section~\ref{sec:preliminary} provides background knowledge.
Section~\ref{sec:methodology} introduces the proposed DPMM-RUL framework.
Section~\ref{sec:simulation} presents simulation studies to evaluate the accuracy, robustness, and sensitivity of the proposed method under various scenarios.
Section~\ref{sec:casestudy} presents a case study of aircraft gas turbine engines to further validate the proposed method. 
Finally, Section~\ref{sec:conclusion} concludes this paper and outlines future research directions.

\section{Literature Review}
\label{sec:literature}
Research on system failure mode identification and prognostics has advanced substantially over the past decades. Early studies \citep{liu2013data, bian2014stochastic} treated failure mode identification and prognostics as separate tasks, and most prognostics models were developed assuming a single failure mode. As manufacturing systems grew more complex, recent studies began to recognize multiple failure modes, each governed by distinct physical degradation processes. This has driven a significant methodological change toward more integrated failure mode identification-prognostics modeling. In the remainder of this section, we will review this evolution and highlight key milestones in failure mode identification and prognostics.

\subsection{Prognostics without Failure Mode Identification}

Prognostic models are generally divided into two main categories: physics-based and data-driven approaches. Physics-based approaches describe the degradation process of systems using mathematical models derived from physical laws. While theoretically grounded, these models often demand a deep understanding of system-specific degradation behaviors, which can be difficult to obtain for increasingly complex modern equipment. More detailed reviews of physics-based models can be found in \cite{soleimani2021diagnostics}. In contrast, data-driven approaches learn patterns from historical and real-time sensor data to predict RUL. They often come with a low implementation cost, especially when detailed physical modeling is impractical or infeasible \citep{vachtsevanos2006intelligent}.

Among various data-driven methods, covariate-based hazard models, pioneered by Cox’s seminal work \citep{cox1972regression}, have been widely applied for external failure factor analysis in various domains \citep{love1991application, jardine1997optimal, lugtigheid2008finite}. For instance, \cite{huh2024integrated} proposed a joint framework that models multi-sensor degradation signals using functional principal component analysis (FPCA) and characterizes time-to-event data through a Bayesian Cox model. 

Recently, a variety of neural network architectures have been explored in prognostics, including feed-forward neural networks (FFNNs), dynamic wavelet neural networks (DWNNs), and self-organizing maps (SOMs). Among them, FFNN is the most widely used due to its simplicity. For example, \cite{wu2007neural} implemented a one-hidden-layer FFNN trained via the Levenberg-Marquardt algorithm to predict the lifetime percentage of bearings using real-time condition monitoring data, which included moving average degradation signals and operational time under worn conditions. \cite{qiu2006wavelet} employed wavelet filtering to denoise vibration signals and used SOMs to detect faults and estimate RUL. 

To better manage the high dimensionality and complexity of modern sensor data, deep learning techniques have been introduced to extract meaningful representations. For instance, \cite{sateesh2016deep} and \cite{li2022deep} applied deep convolutional neural networks (CNNs) to extract temporal features from sequential sensor inputs. Furthermore, recurrent neural networks (RNNs) \citep{guo2017recurrent}, including their advanced forms such as Long Short-Term Memory (LSTM) \citep{huang2019bidirectional, kim2020bayesian} and Gated Recurrent Units (GRU) \citep{ni2022data}, have been widely adopted to model temporal dependencies and enhance the accuracy of RUL predictions.

\subsection{Prognostics with Failure Mode Identification} 

As both systems and their operating conditions become complicated, research specifically combining failure mode identification with prognostics has gained attention. In the literature, two dominant modeling paradigms have emerged based on the physical characteristics of the system. 

The first approach is the competing risks approach \citep{Wang2012j, Song2016q, Jiang2015v, Rafiee2014l, Peng2011k, Fan2019j}. In this approach, each system is subject to several potential failure processes operating in parallel. The system fails when the earliest of these processes reaches its failure threshold. This scenario corresponds to a serial system, where the failure of any component results in overall system failure. Under common assumptions (e.g., independent failure processes), system reliability can be derived from the individual reliabilities of each failure mode.

The second approach involves mixture models, where each system corresponds to a specific failure mode \citep{chehade2018, song2019multi, wang2024, fu2025degradation}. These models represent the population or its degradation trajectories as a mixture of latent, failure mode-specific behaviors. Each system is governed by a single dominant failure mode (DFM) throughout its lifecycle. Rather than multiple mechanisms competing simultaneously, one mechanism dominates the degradation process from the outset. For example, in bearings, once a crack initiates, stress concentration at the crack tip accelerates propagation far faster than competing mechanisms like general corrosion. The crack becomes the DFM, while other processes remain secondary throughout the bearing's life. The proposed method also belongs to this second approach. 

Within the mixture model framework, several methods have been developed. \cite{chehade2018} developed a composite failure-mode index through multi-sensor data fusion, enabling online estimation of mode-specific probabilities and subsequent polynomial degradation modeling. \cite{song2019multi} extended this approach by sequentially addressing failure mode identification and RUL prediction using separate neural network branches. \cite{wang2024} further improved this by simultaneously learning failure mode identification and RUL prediction through a joint deep learning model, significantly enhancing overall predictive accuracy. \cite{fu2025degradation} removed the requirement for failure mode labels, but still required knowing the exact and fixed number of failure modes to cluster degradation trajectories inferred from low-dimensional manifold representations.

Despite these advances, current approaches are facing several limitations. First, many methods separately handle failure mode identification and RUL prognostics, potentially missing shared information between these closely related tasks \citep{wang2024deep, chen2022wafer, ramasso2014performance}. Second, these methods often assume a fixed and known number of failure modes, which can limit their adaptability to non-stationary and evolving environments where new failure modes may continuously appear. Third, they typically presume that the failure modes of historical (failed) systems are known. However, in practice, failed systems are often discarded without identifying their primary failure mechanisms. These limitations highlight the need for more robust and adaptive approaches that can jointly learn failure mode and prognostic representations, autonomously uncover the underlying failure mechanisms, and dynamically adapt to emerging mechanisms in real-world manufacturing systems.

\section{Preliminary - Dirichlet Process and Stick-Breaking}
\label{sec:preliminary}

The Dirichlet process (DP) is a stochastic process over probability measures \citep{teh2017dirichlet}. It is parameterized by a concentration parameter \( \alpha \) and a base distribution \( H \). A random probability measure \( G \) is said to follow a DP, denoted as:
\begin{equation}
    G \sim \text{DP}(\alpha, H),
\end{equation}
when for any finite measurable partition \( \{A_1, \dots, A_k\} \), the random vector \( (G(A_1), \dots, G(A_k)) \) follows a Dirichlet distribution with parameters \( (\alpha H(A_1), \dots, \alpha H(A_k)) \). Here, $H(A_i)$ denotes the probability mass that base distribution $H$ assigns to the measurable set $A_i$, and $G(A_i)$ is the corresponding random probability mass assigned by the random measure $G$.

\begin{figure*}[!t]
\centering
\includegraphics[width=2.5in]{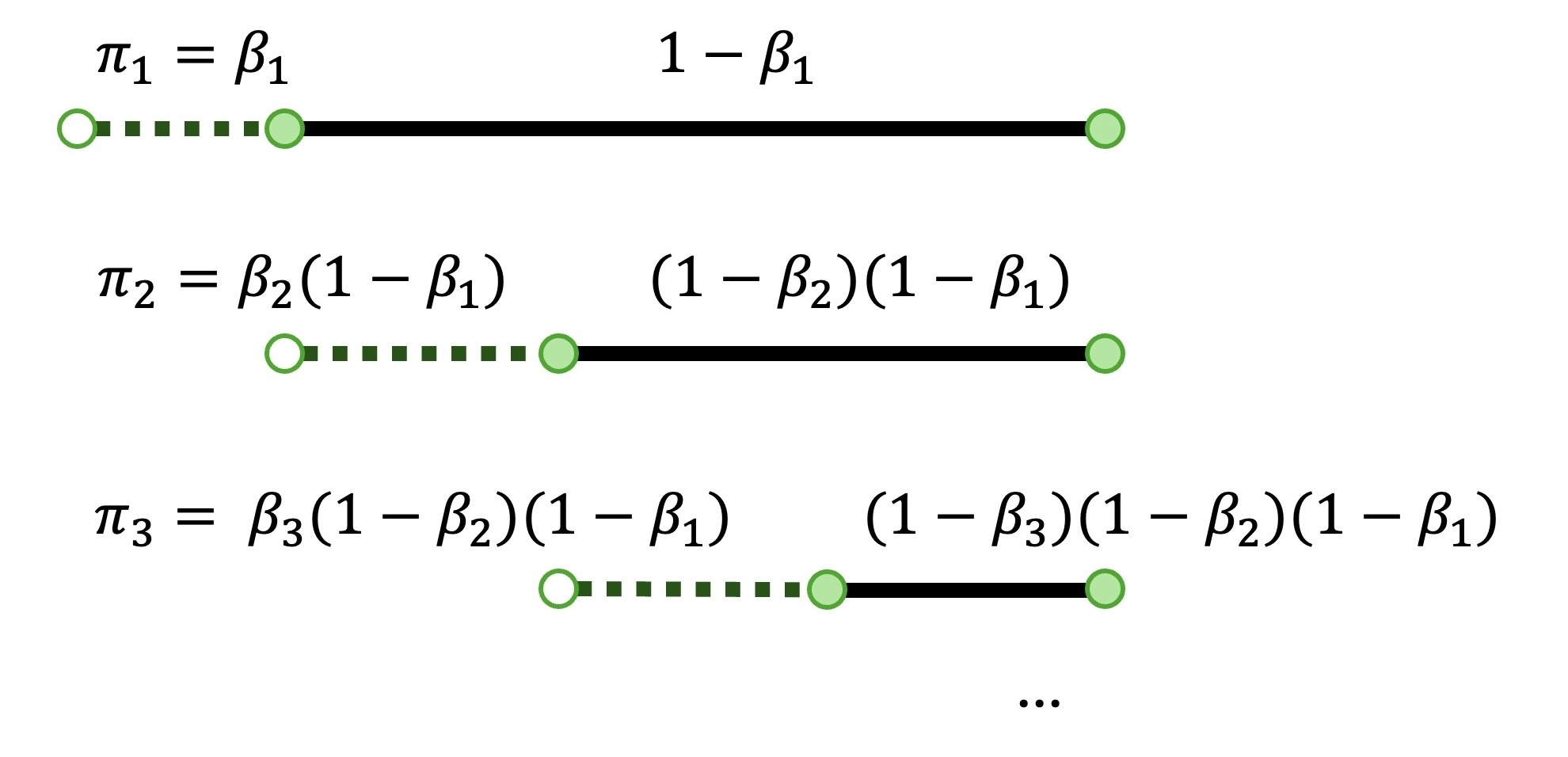}%
\includegraphics[width=2.5in]
{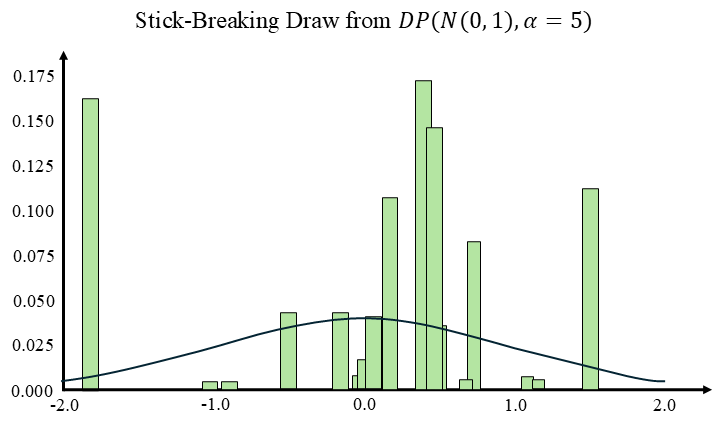}%
\caption{ (a) Stick-breaking process. (b) A sample draw from a DP.}
\label{fig:stick_breaking}
\end{figure*}

For constructive purposes, the DP can be represented using the stick-breaking process, as illustrated in Figure~\ref{fig:stick_breaking}. This process can be interpreted as sequentially breaking a unit-length stick into an infinite number of segments, where each segment $k$ represents the weight \( \pi_k \) assigned to component \( k \). Specifically, for \( k \geq 1 \), the mixture weights \( \pi_k \) are defined as:
\begin{equation}
\label{eq:stick-breaking}
    \pi_k = \beta_k \prod_{i=1}^{k-1} (1 - \beta_i), \quad \text{with } \beta_k \sim \text{Beta}(1, \alpha).
\end{equation}
As the process continues, the weights satisfy \( \sum_{k=1}^{\infty} \pi_k = 1 \) and the resulting distribution over \( \boldsymbol{\pi} = \{\pi_k\}_{k=1}^\infty \) is known as the Griffiths-Engen-McCloskey (GEM) distribution:
\begin{equation}
    \boldsymbol{\pi} \mid \alpha \sim \text{GEM}(\alpha).
\end{equation}
To complete the construction of a random probability measure, atoms $\theta_k$ are drawn independently from the base distribution $H$. The random measure $G$ is then expressed as a weighted sum of Dirac delta measures:
\begin{equation}
    G(A)=\sum_{k=1}^\infty \pi_k \delta_{\theta_k}.
\end{equation}
Figure~\ref{fig:stick_breaking} (b) shows a sample draw (a random distribution) from a DP with base distribution $H=N(0,1)$ and concentration parameter $\alpha=5$.
The DP serves as a prior distribution in the DPMM, which will be further explained in Section \ref{sec:DPMM}.

\section{Proposed Method}
\label{sec:methodology}

This section presents our proposed methodology for addressing the two main challenges of failure mode identification and prognostics in complex and dynamic environments. Specifically, we aim to (i) automatically infer latent failure modes from multi-sensor data, and (ii) accurately predict the RUL of the system.

\begin{figure}[!bt]
    \centering
    \includegraphics[width=5.5in]{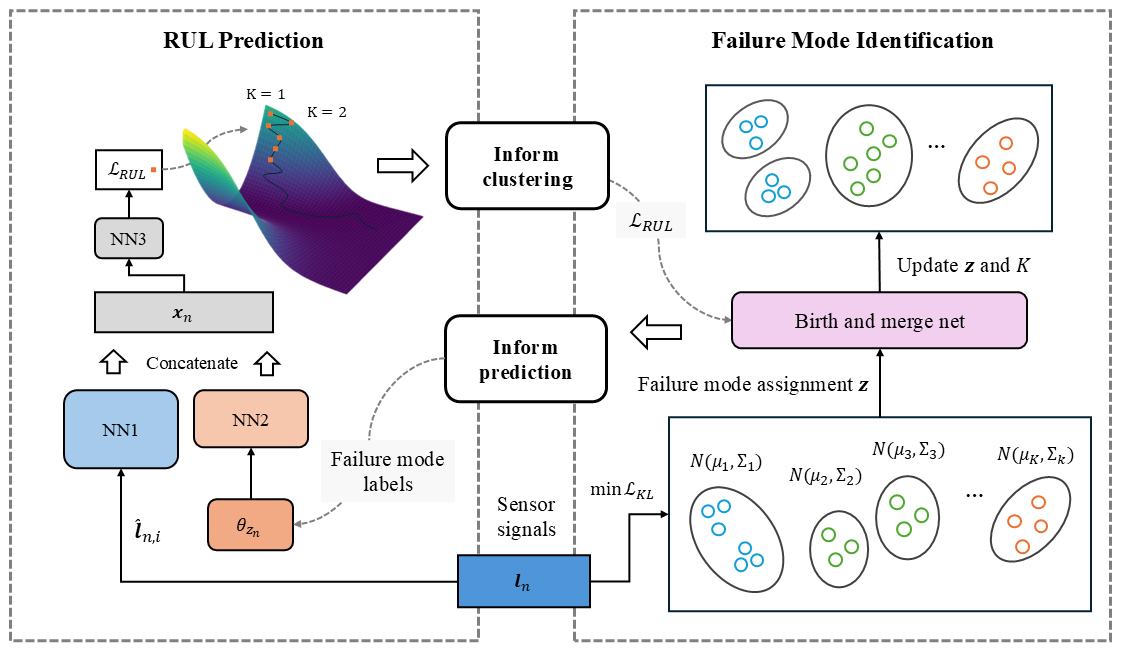}
    \caption{Overview of the DPMM-RUL}
    \label{fig:DPMM-RUL}
\end{figure}

An overview of the proposed framework is provided in Figure~\ref{fig:DPMM-RUL}. The proposed framework consists of two modules: a DPMM-based failure mode identification module and an RUL prediction module. 
Section~\ref{sec:DPMM} introduces the Bayesian nonparametric modeling of the failure mode identification module. Section~\ref{sec:Prognostics_NN} introduces the failure mode informed RUL prognostics module. Section~\ref{sec:inference_and_update} outlines the parameter inference and update of both modules. Section~\ref{sec:online_VI} discusses the iterative update between the two modules through the online variational inference.

\subsection{Nonparametric Modeling of Degradation Data under Unknown and Unlabeled Failure Modes} 
\label{sec:DPMM}
We begin by formulating the failure mode identification problem. Consider $N$ systems whose degradation behaviors are governed by multiple, distinct but unknown failure mechanisms. Each system is monitored by $S$ sensors. The functional relationship between these multi-sensor signals and the system's RUL is unknown and potentially nonlinear and varies depending on its failure mode.
Neither the types nor the number of failure modes is known in advance, and thus a modeling framework is required to automatically discover and characterize these modes from data.

For system $n$, let $l_{n, s, t}$ denote the reading of sensor $s \in \{1,\ldots,S\}$ at time $t$. Since different systems operate for different durations, the total number of time steps (e.g., operating cycles) $T_n$ may vary across systems. To manage the varying operational durations $T_n$ across systems, for each system $n$, its multi-sensor signals are converted to a representation with fixed-length $\bar T$. Multiple strategies have been proposed in literature to achieve this, including truncation and padding, temporal warping and summary feature extraction. 

For instance, truncation and padding set a reference length $\bar{T}$ (e.g., the mean duration across systems $\bar{T}=\frac{1}{N} \sum_{n=1}^N T_n$). If $T_n>\bar{T}$, the sequence retains the last (most recent) $\bar{T}$ cycles.  If $T_n\leq\bar{T}$, the sequence is padded to length $\bar{T}$ using the last observed sensor signals. This preserves the most recent degradation behavior prior to failure. The choice of preprocessing depends on domain knowledge and users' needs, e.g., whether absolute time-to-failure or degradation trajectory shape plays a primary role in failure mode identification. In the numerical studies, we further investigate the effect of different preprocessing strategies on inferred failure modes and prognostic performance.
The resulting complete dataset of the multivariate sensor signals is represented as:
$$
\mathbf{l} = \{\mathbf{l}_n\}_{n=1}^N,
$$
where $\mathbf{l}_n$ is signals from system $n$ and can be flattened into a vector of fixed dimension $D=S\times \bar{T}$.

The goal of our failure mode identification module is to infer the latent failure mode from these collected sensor signals. A natural approach is to model the heterogeneous population as a finite mixture:
\begin{equation}
p(\mathbf{l}_n|\mathbf{\theta},\mathbf{\pi})=\sum_{k=1}^K\pi_k F(\mathbf{l}_n|\theta_k),
\end{equation}
where $K$ is the number of failure modes, $\mathbf{\pi}=\{\pi_1,...,\pi_K\}$ are mixture weights, $\mathbf{\theta}=\{\theta_1,...,\theta_K\}$ are failure mode-specific parameters, and $F(\cdot|\theta_k)$ is the likelihood of sensor readings under failure mode $k$. In other words, each failure mode $k$ is governed by parameters $\theta_k$ that determine the sensor data distribution. This mixture model, in fact, has been widely used in multi-failure mode prognostic studies \citep{jiang2016gmm}.

However, a critical limitation is that the number of failure modes $K$ must be fixed and specified in advance. In real-world prognostic applications, where underlying failure mechanisms are uncertain or evolving, misspecifying $K$ can lead to poor RUL predictions. It is therefore critical to treat $K$ as an unknown quantity and infer it jointly with the model parameters from data.

To overcome this limitation, we formulate a hierarchical Bayesian nonparametric model based on the DP prior. The DP prior naturally induces a countably infinite set of potential clusters (failure modes), thereby allowing the data to automatically determine how many failure modes are actually needed. This yields a flexible mixture model in which multiple systems can share the same cluster assignment, and new clusters can be created automatically if justified by the data.

Building on the DP construction introduced in Section \ref{sec:preliminary}, let $\{\theta_k\}_{k=1}^\infty$ denote the infinite collection of potential failure mode-specific parameters generated from a base distribution $H$ with hyperparameter $\lambda$. The infinite-dimensional mixture weight vector $\mathbf{\pi}$ is defined by stick-breaking variables $\{\beta_k\}_{k=1}^\infty$ with concentration parameter $\alpha$.

Each system $n$ is assigned a latent failure mode indicator $z_n\in\{1,2,...\}$ drawn from the weights $\mathbf{\pi}$. The observed sensor sequence $\mathbf{l}_n$ for system $n$ is then generated according to a likelihood model $F$ conditioned on the specific failure mode parameters $\theta_{z_n}$. The resulting hierarchical model is defined as follows:
\begin{equation}
\begin{aligned}
    \theta_k \mid H &\sim H(\lambda), \\
    \beta_k \mid \alpha &\sim \text{Beta}(1, \alpha), \\
    z_n \mid \{\beta_k\}_{k=1}^\infty &\sim \text{Cat}(\mathbf{\pi(\beta)}), \text{ and} \\
    \mathbf{l}_n\mid \theta_{z_n} &\sim F(\mathbf{l}_n|\theta_{z_n}).
\end{aligned}
\label{eq:draw_from_DPMM}
\end{equation}

\begin{figure}[tb]
    \centering
    \includegraphics[width=0.5\linewidth]{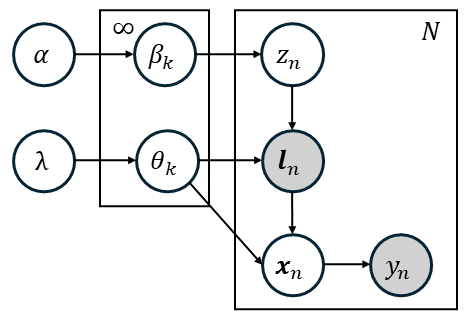}
    \caption{Graphical representation of the proposed framework. Shaded nodes indicate observed variables, while unshaded nodes indicate latent (unobserved) variables or parameters.}
    \label{fig:graphical_diagram}
\end{figure}

The graphical model representation of the proposed hierarchical model is shown in Figure~\ref{fig:graphical_diagram}.
The observed signals $\mathbf{l}_n$ and underlying parameters $\theta_k$ are then embedded into a latent representation $\mathbf{x}_n$. The prognostic module, which will be explained in the next subsection, maps this latent representation to the system's RUL $y_n$ through a deep learning function $\phi_g$. 

The probabilistic failure mode assignments inferred by the DPMM provide a cluster representation. However, failure mode identification alone does not directly yield system prognostics such as RUL. In the next subsection, we introduce a neural network-based prognostics model that explicitly conditions RUL prediction on the inferred failure mode information.

It is important to note that, unlike existing DPMM-based models that employ mixture components only to capture different contextual information (e.g., \cite{kim2023}), the proposed framework applies DPMM to jointly solve two main challenges in system prognostics simultaneously. As a result, the role of the DPMM in our framework extends beyond density modeling. It serves as one component within a mutually informed learning architecture.

\subsection{Failure Mode-Informed RUL Prognostics}
    \label{sec:Prognostics_NN}
    
Building on the failure modes inferred by the DPMM in Section \ref{sec:DPMM}, we design a RUL prognostic module that incorporates both the observable sensor signal and the latent failure modes. Rather than directly mapping sensor signals to RUL, the proposed module adaptively adjusts its predictions based on the inferred failure mode.

Given the sensor signal data, training samples are constructed using a sliding window of length $\xi$. Specifically, the $i$th window extracted from system $n$, denoted by $\hat{\mathbf{l}}_{n, i}\in\mathbb{R}^{S\times\xi}$, is defined as: 
$$
\hat{\mathbf{l}}_{n,i}=\{\mathbf{l}_{n,t}\}_{t=i}^{i+\xi-1},
$$
where $\mathbf{l}_{n,t} = (l_{n,1,t}, \ldots, l_{n,S,t})^\top \in \mathbb{R}^{S}$ and $i\in\{1,...,T_n-\xi+1\}$. This results in a total of
$
\sum_{n=1}^N (T_n-\xi+1)
$
training samples. The RUL corresponding to the $i$th sample $\hat{\mathbf{l}}_{n,i}$ is then $y_{n,i}=T_n-(i+\xi-1)$.

The prognostic module processes each sample through three neural networks with distinct roles. The first is a signal encoding network $\phi_l: \mathbb{R}^{S\times\xi} \rightarrow \mathbb{R}^{d_u}$, transforming the signal data $\hat{\mathbf{l}}_{n,i}$ into a latent signal embedding $\phi_l(\hat{\mathbf{l}}_{n,i})\in\mathbb{R}^{d_u}$. The second is a failure-mode context network $\phi_\theta: \mathbb{R}^{d_\theta} \rightarrow \mathbb{R}^{d_v}$ that maps the failure mode parameter $\theta_{z_n}\in\mathbb{R}^{d_\theta}$ to a compact latent context embedding $\phi_\theta(\theta_{z_n})\in\mathbb{R}^{d_v}$. 

These two latent representations are concatenated to form a joint feature vector:
\begin{equation}
\label{eq:rul_feature}
    \mathbf{x}_{n,i} = [\phi_l(\hat{\mathbf{l}}_{n,i}); \phi_\theta(\theta_{z_n})],
\end{equation}
which is then processed by a prediction network $\phi_g:\mathbb{R}^{d_u+d_v} \rightarrow \mathbb{R}$ to estimate the RUL:
$$
\hat{y}_{n,i} = \phi_g(\mathbf{x}_{n,i}).
$$
The proposed prognostic module implicitly adapts to different degradation behaviors under different failure modes.

\begin{figure}[!t]
\centering
\includegraphics[width=5.5in]{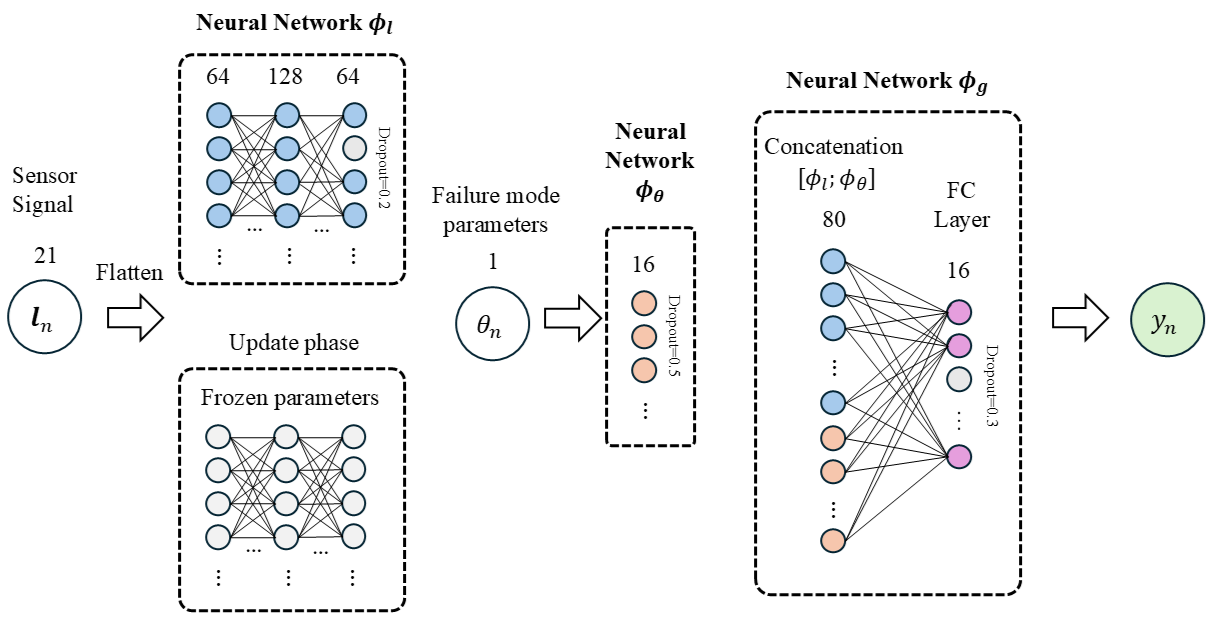}
\caption{Neural network structure of RUL prediction model}
\label{fig:RUL_NN_structure}
\end{figure}

The proposed framework is not restricted to a specific network architecture. Various neural network structures capable of modeling temporal sensor data may be employed without altering the underlying formulation. In Sections \ref{sec:simulation} and \ref{sec:casestudy}, standard multilayer perceptrons (MLPs) are adopted for simplicity and efficiency as illustrated in Figure \ref{fig:RUL_NN_structure}. The signal model $\phi_l$ is implemented as a three-layer MLP yielding a 64-dimensional embedding ($d_u=64$). The context model $\phi_\theta$ is a single-layer MLP yielding a 16-dimensional embedding ($d_v=16$). These representations are concatenated into an 80-dimensional feature vector $\mathbf{x}_{n,i}$.

Given the failure mode identification and prognostic modules, the next step is to estimate their parameters. In the following subsection, we describe the inference and update procedures for each module, which serve as the foundation for the iterative learning introduced in Section \ref{sec:online_VI}.

\subsection{Inference and Parameter Estimation}

This section describes the inference and parameter estimation procedures for the failure mode identification and prognostics modules. 
It is worth mentioning that, intuitively, we can model a joint distribution including all variables $p(\mathbf{l}, y, z, \theta, \beta)$ and conduct inference. However, the high-dimensional $\mathbf{l}$ overshadows the RUL score $y$, and this standard approach results in significantly high RUL prediction errors. This joint inference approach will be studied as a benchmark in Sections~\ref{sec:simulation} and \ref{sec:casestudy}. To address this, we present the variational inference for the DPMM assuming fixed prognostic parameters, and the training of the prognostics model given fixed and updated failure mode assignments.

\label{sec:inference_and_update}
\subsubsection{Variational Inference of Failure Mode Identification Module}
\label{ssec:inference}

The DPMM-based failure mode identification module updates the posterior over the latent failure mode assignments $z=\{z_n\}_{n=1}^N$, the failure mode parameters $\theta = \{\theta_k\}_{k=1}^{\infty}$ and the stick-breaking variables $\beta = \{\beta_k\}_{k=1}^{\infty}$. As described in Section \ref{sec:DPMM}, sensor signals from each system are aligned to a fixed length $\bar{T}$, resulting in a fixed-dimensional observation $\mathbf{l}_n\in\mathbb{R}^D$. 
The joint distribution can be written as:

\begin{equation}
    p(\mathbf{l}, z, \theta, \beta) = \prod_{n=1}^N \ F(\mathbf{l}_n \mid\theta_{z_n}) \text{Cat}(z_n \mid \boldsymbol{\pi}) \prod_{k=1}^{\infty} \text{Beta}(\beta_k \mid 1, \alpha) H(\theta_k \mid \lambda),
\end{equation}
where \(\boldsymbol{\pi}\) is constructed via the stick-breaking process, i.e., \(\pi_k = \beta_k \prod_{j=1}^{k-1}(1-\beta_j)\). 

We assume that $F$ is a multivariate Gaussian distribution and $H$ is its conjugate Normal-Inverse-Wishart (NIW) distribution:
\begin{equation}
    F(\mathbf{l}_n \mid \theta_k) = \mathcal{N}(\mathbf{l}_n \mid \boldsymbol{\mu}_{\theta_k}, \boldsymbol{\Sigma}_{\theta_k}),
\end{equation}
\begin{equation}
    H(\theta_k \mid \lambda) = \mathcal{NIW}(\boldsymbol{\mu}_{\theta_k}, \boldsymbol{\Sigma}_{\theta_k} \mid \lambda),
\end{equation}
where $\theta_k = \{\boldsymbol{\mu}_{\theta_k}, \boldsymbol{\Sigma}_{\theta_k}\}$ denotes the mean vector and diagonal covariance matrix characterizing the $k$-th failure mode and $\lambda=\{\mathbf{m}_0, \kappa_0, \nu_0, \boldsymbol{\Psi}_0\}$ are the hyperparameters of the base distribution.

Since the exact posterior $p(z,\theta,\beta\mid \mathbf{l})$ is intractable, we resort to a mean-field variational inference by introducing a tractable variational distribution $q( z, \theta, \beta)$ that is fully factorized across latent variables and parameters:
\begin{equation}
\label{eq:mean-field}
q(z,\theta,\beta) =
\prod_{n=1}^N q(z_n)
\prod_{k=1}^{K-1} q(\pi_k)
 \prod_{k=1}^{K} q(\theta_k),
\end{equation}
with:
\begin{align}
q(z_n) &= \text{Cat}(z_n \mid \hat{r}_{n1}, \dots, \hat{r}_{nK}), \\
q(\beta_k) &= \text{Beta}(\beta_k \mid \hat{\alpha}_{0,k}, \hat{\alpha}_{1,k}), \\
q(\theta_k) &= \mathcal{NIW}(\theta_k \mid \hat{\lambda}_k).
\end{align}
where $\hat{r}_{n}=\{\hat{r}_{nk}\}_{k=1}^{K}$ is the parameter of a categorical distribution and represents the responsibility of failure modes for the $n$-th system, $\hat{\alpha}_{0, k}$ and $\hat{\alpha}_{1, k}$ are parameters for Beta distribution, and $\hat{\lambda}_k$ is the variational posterior parameters for failure mode $k$, i.e., $\hat{\lambda}_k=\{\mathbf{m}_k, \kappa_k, \nu_k, \boldsymbol{\Psi}_k\}$. 

To ensure tractability, the variational distribution is truncated at a fixed level $K$. It is important to note that this truncation only applies to the variational approximation, not the underlying DP prior \citep{blei2006}. A sufficiently large $K$ allows the variational posterior to capture all necessary failure modes without fully utilizing all $K$ components. The DP prior promotes sparsity by assigning negligible weights to redundant components. More importantly, the proposed method introduces a birth-merge mechanism (Section \ref{sec:online_VI}) to further enable the model to automatically increase or decrease the effective number of active failure modes during training. Under the truncation, the final mixture weight is set to ensure that the mixture weights sum to 1.

Our goal is to minimize the Kullback-Leibler (KL) divergence to the true posterior:
\begin{equation}
q^*(z,\theta,\beta) = \arg\min_{q} \,
\mathrm{KL}\bigl(q(z,\theta,\beta) \,\mid\, p(z,\theta,\beta \mid \mathbf{l})\bigr).
\end{equation}
Instead of directly minimizing the KL divergence, we equivalently maximize the Evidence Lower Bound (ELBO), which provides a tractable objective function for optimization:
\begin{align}
\mathcal{L}_{\text{ELBO}}(q)
    &= \mathbb{E}[\log p(\mathbf{l}, z, \theta, \beta)]
    - \mathbb{E}[\log q(z,\theta,\beta)] \notag\\
  &= \mathbb{E}[\log p(z, \theta, \beta)]
     + \mathbb{E}[\log p(\mathbf{l} \mid z, \theta, \beta)]
     - \mathbb{E}[\log q(z, \theta, \beta)] \notag\\
  &= \mathbb{E}[\log p(\mathbf{l} \mid z, \theta, \beta)]
     - \mathrm{KL}\bigl(q(z,\theta,\beta) \mid p(z,\theta,\beta)\bigr)
     \label{eq:elbo}
\end{align}
The first term, $\mathbb{E}[\log p(\mathbf{l} \mid z, \theta, \beta)]$, represents the expected log-likelihood under the variational distribution, which encourages parameter configurations that accurately model the observed signal data. The remaining terms constitute the KL divergence $\text{KL}(q(z,\theta,\beta) | p(z,\theta,\beta))$, which regularizes the variational posterior toward the prior distributions, thereby preventing overfitting and ensuring proper Bayesian inference.

Under the mean-field assumption, Equation~\eqref{eq:elbo} can be rewritten as:
\begin{align}
\mathcal{L}_{\text{ELBO}}(q)
&= \mathbb{E}_q[\log p(\mathbf{l} \mid \theta,z)]
+ \mathbb{E}_q[\log p(z \mid \beta)]
+ \mathbb{E}_q[\log p(\beta \mid \alpha)]
+ \mathbb{E}_q[\log p(\theta \mid \lambda)] \notag \\
&\quad - \mathbb{E}_q[\log q(z)]
- \mathbb{E}_q[\log q(\beta)]
- \mathbb{E}_q[\log q(\theta)].
\label{eq:elbo-expanded}
\end{align}

To optimize the ELBO in Equation~\eqref{eq:elbo-expanded}, we adopt coordinate ascent variational inference (CAVI). For the $j$th factor \( q^*(\psi_j) \), the optimal update is:
\begin{equation}
    q^*(\psi_j) \propto \exp\left( \mathbb{E}_{q_{-j}}[\log p(\psi_j, \psi_{-j}, \mathbf{l})] \right),
\end{equation}
where the logarithm is taken over the joint distribution of all latent variables and observations, and the expectation is taken over all latent variables except \( \psi_j \). This allows for tractable updates when the joint distribution belongs to the exponential family, and the dependencies among latent variables are simplified via the mean-field assumption. The full equations for updating the variational parameters in the algorithm are presented in the Appendices.

\subsubsection{Training of Prognostics Module}

We now describe the training procedure for the prognostics module. Given the current failure mode identification $z$ and parameters $\theta$, the failure mode context embedding $\phi_\theta(\theta_{z_n})$ is fully determined for each system $n$. Together with the signal embedding $\phi_l(\hat{\mathbf{l}}_{n,i})$, this defines the feature vector $\mathbf{x}_{n,i}$ needed for RUL prediction. Let $\Phi$ denote the learnable parameters of the signal embedding, the failure-mode context embedding, and the fusion networks in the prognostics module. The module is trained by minimizing the Root Mean Square Error (RMSE) between predicted and true RUL values:
\begin{equation}
    \mathcal{L}_{\text{RUL}}(\Phi) = \sqrt{\frac{1}{\sum_{n=1}^N (T_n-\xi+1)}\sum_{n=1}^N\sum_{i=1}^{T_n-\xi+1} \left( y_{n,i} - \phi_g([\phi_l(\hat{\mathbf{l}}_{n,i}); \phi_\theta(\theta_{z_n})]) \right)^2}.
\end{equation}

In practical deployment settings, data arrive sequentially and unobserved failure modes may appear. To accommodate such non-stationarity, the proposed learning procedure is therefore fully iterative, with model parameters updated continuously as new data become available. At the initial stage, all parameters are pretrained jointly  for a limited number of epochs through gradient descent:
\begin{equation}
    \Phi \leftarrow \Phi - r \nabla_{\Phi} \mathcal{L}_{\text{RUL}},
\end{equation}
where $r$ is the learning rate.

After this initial pre-training phase, the parameters of the signal embedding network $\phi_l$ are fixed. Empirically, we observed that repeatedly updating these parameters in subsequent iterations yields little performance improvement while incurring unnecessary computational costs. 
In contrast, the failure mode context embedding network $\phi_\theta$ and fusion network $\phi_g$ are updated continuously, as they must adapt to the constantly refined failure mode identification produced by the DPMM module.
Accordingly, the iterative updates are given by
\begin{equation}
    \varphi_\theta \leftarrow \varphi_\theta - r \nabla_{\varphi_\theta} \mathcal{L}_{\text{RUL}} \quad \text{and} \quad \varphi_g \leftarrow \varphi_g - r \nabla_{\varphi_g} \mathcal{L}_{\text{RUL}},
\end{equation}
where $\varphi_\cdot$ denotes the trainable parameters of network $\phi_\cdot$.

As the training proceeds, the number of inferred failure modes may change. Once the number of failure modes stabilizes over a sustained period, the model treats the current structure as locally converged. If a failure mode structure change is detected, such as the addition or removal of a failure mode, the signal embedding parameters $\varphi_l$ are updated to accommodate such changes. This selective update strategy ensures the model preserves stable signal representations when the environment is stationary, while retaining flexibility to refine when the failure mode behaviors change.

\subsection{Online Variational Inference for Interaction between Two Modules}
\label{sec:online_VI}
We perform online learning by iteratively updating the failure-mode identification module and the prognostic module. Although traditional variational inference provides a principled approach for model updates, optimizing solely the ELBO can easily converge to locally optimal configurations due to the Dirichlet process ``rich-get-richer'' effect.

To address these limitations, we propose an online variational inference framework that explicitly incorporates prognostic feedback into the structural learning of failure modes. Specifically, we define a DPMM-RUL performance score, $\mathcal{J}$,
to guide these structural changes by coupling failure mode identification with prognostic accuracy. 
The DPMM-RUL performance score is defined as:
\begin{equation}
\label{eq: DPMM-RUL performance}
    \mathcal{J}(\mathbf{z},\phi)
    \;=\;
     \operatorname{Sil}(z;\mathbf{l}) - \omega \cdot \mathcal{L}_{\text{RUL}}(\phi), 
\end{equation}
where $\operatorname{Sil}(z;\mathbf{l})$ is the Silhouette Index, serving as a geometric regularizer that encourages failure modes to be internally compact and externally well-separated in the signal space. $\mathcal{L}_{\text{RUL}}(\phi)$ is the RUL prediction loss, ensuring the failure mode identification is informed by the prognostic result, and $\omega>0$ is the trade-off coefficient tuned using K-fold cross-validation to balance failure mode identification performance against prediction accuracy. 

Under this framework, the proposed learning procedure seeks to increase $\mathcal{J}$. After each round of updating the DPMM and RUL components, we apply birth and merge moves, which adaptively modify the number of active failure modes $K$ during training, and make a structural change (i.e., a birth or merge move) if $\Delta\mathcal{J}\ge 0$. The efficiency of this score is further demonstrated in the ablation study in Section~\ref{sec:simulation}.

\subsubsection{Birth Moves to Escape Local Optima}
\hspace{0.5em}
The birth move provides a mechanism for exploring new possible failure modes that may not be reachable through standard coordinate-ascent updates. We develop an online adaptation of the birth strategy inspired by \cite{hughes2013memoized}. For a selected failure mode $k'$, a subset of systems with responsibility $\hat{r}_{nk'} \ge r_{threshold}$ (e.g., $r_{threshold} = 0.1$) are sampled. A separate DPMM with a small truncation level $K'$ is then fitted to this subset via variational inference for a limited number of iterations. From our simulation and case studies, we observed that the proposed method works well with a small value of $K'=1,2,3$.

The resulting local mixture components are tentatively added to the global mixture, expanding the current mixture from truncation level $K$ to $K+K'$. System-level failure mode assignments are recomputed based on the updated mixture model. At this stage, we do not evaluate the ELBO change associated with the new components. All births are initially accepted. Components that are empty, redundant, or overly similar to existing modes will be removed through the subsequent merge operation.

\subsubsection{Merge Moves to Jointly Optimize Two Modules}
\hspace{0.5em}
The birth moves help the algorithm to escape from local optima. The merge move, on the other hand, serves as a mechanism by which prognostic feedback influences the failure mode structure. In the absence of prognostic information, the birth move might retain redundant failure modes that model noise rather than meaningful failure modes.

To prevent this, we evaluate candidate pairs of failure modes for merging. Unlike standard approaches that rely solely on the ELBO \citep{hughes2013memoized}, we employ the joint performance criterion $\mathcal{J}$ to assess candidate merges. For each candidate pair, we simulate the merge, update the affected model components, and recompute $\mathcal{J}$. A merge is accepted only if it leads to an improvement in the performance score. Intuitively, if distinguishing between two failure modes does not improve RUL prediction accuracy, treating them as separate modes is unnecessary.

The birth and merge moves help further adjust the failure mode identification to ensure the final model structure is determined not just by sensor signal statistics as in approaches that identify failure modes only based on sensor data (e.g., \cite{fu2025degradation}), but by the joint maximization of failure mode identification performance and RUL prediction accuracy without sacrificing interpretability. The complete birth-merge algorithm is summarized in Algorithm~\ref{alg:birth and merge}.

\begin{algorithm}[H]
\caption{The birth and merge moves of DPMM-RUL}
\label{alg:birth and merge}
\begin{algorithmic}[1]
\Require Sensor signal dataset $\mathbf{l}$, sample threshold $r_{threshold}$, current failure mode assignments $z_n$, new truncation level $K'$
\Ensure updated failure mode assignments $z'_n$
\State \textbf{Birth:}
    \State \quad Randomly select a failure mode $k'$
    \State \quad Sample a subset  $\mathbf{l}' \subset \mathbf{l}$ by selecting systems whose responsibility $\hat{r}_{nk'} \ge r_{threshold}$; fit local DPMM with $K'$ initial failure modes
    \State \quad Tentatively add new failure modes to the global model
    \State \quad Reassign all $z'_n$ over updated failure mode set
\State \textbf{Merge:}
    \State \quad For each failure mode pair $(k_a,k_b)$, where $k_a, k_b\in \{1,...,K\}$ and $k_a\neq k_b$, evaluate merge-candidate's $\mathcal{J}$
    \State \quad Reject the merge if $\Delta\mathcal{J}\leq 0$. Otherwise, accept the merge with the largest $\Delta\mathcal{J}$
    \State \quad Reassign all $z'_n$ over updated failure mode set
\end{algorithmic}
\end{algorithm}

The complete DPMM–RUL framework alternates between two stages. First, variational inference is used as an inner optimization step to estimate latent failure mode assignments and parameters for a fixed model structure. Second, birth and merge operations adapt the model structure using the RUL-informed performance score. The full procedure is summarized in Algorithm~\ref{alg:DPMM-RUL}.

\begin{algorithm}[H]
\caption{DPMM-RUL framework}
\label{alg:DPMM-RUL}
\begin{algorithmic}[1]
\Require Original dataset $\mathcal{D} = \{(\mathbf{l}_n, y_n)\}_{n=1}^N$; initial truncation level $K$; new truncation level $K'$;  maximum iterations $T$
\Ensure Learned number of failure modes, learned parameters of each DPMM failure mode $\beta_{1:K}, \mu_{\theta_{1:K}}, \Sigma_{\theta_{1:K}}$
\State Initialize the DPMM with $K=1$, including variational parameters: $\hat{r}_{nk}, \mu_{\lambda_{1:K}}, \Sigma_{\lambda_{1:K}}$ and Beta-distribution with $\alpha$
\State Initialize RUL prediction network parameters
\State Initialize $\texttt{trained}=\texttt{False}$
\Repeat
    \State \textbf{DPMM Variational Inference:}
    \State Compute responsibilities $\hat{r}_{nk}\propto q(z_n=k\mid\mathbf{l}_n,\hat{\theta},\hat{\beta})$
    \State Update $\{\hat{\theta}_k,\hat{\beta}_k\}$ by maximizing the ELBO, obtaining $\mathcal{L}_\text{ELBO}$
    \State Assign failure mode labels $z_n = \arg\max_k \hat{r}_{nk}$
    \State Compute failure mode identification performance: $\operatorname{Sil}(\{\mathbf{l}_n\},\{z_n\})$

    \State \textbf{RUL Prediction Update:}
    \For{$n=1,\dots,N$}
    \For{$i=1, 2 ..., T_n-\xi+1$}
    \State Mapping signals into embedding $\phi_l(\hat{\mathbf{l}}_{n, i})\quad$
    \State Mapping failure mode parameters into embedding $\phi_\theta(\theta_{z_n})\quad$
    \State Concatenate $\phi_l(\hat{\mathbf{l}}_{n, i})$ and $\phi_\theta(\theta_{z_n})$ into $\mathbf{x}_{n,i}= [\,\phi_l(\hat{\mathbf{l}}_{n, i});\,\phi_\theta(\theta_{z_n})\,]$
  \EndFor
  \EndFor
  \If {$\texttt{trained}==\texttt{False}$} 
  \State Initial pretraining: $\Phi \leftarrow \Phi - r \nabla_{\Phi} \mathcal{L}_{\text{RUL}}$
  \State $\texttt{trained} = \texttt{True}$
  \Else 
    \State Freeze signal embedding parameters $\varphi_l=\varphi_l^{(\text{fixed})}$
    \State Train using current failure mode labels $z_n$:
    \State $\varphi_\theta \leftarrow \varphi_\theta - r \nabla_{\varphi_\theta} \mathcal{L}_{\text{RUL}}, \quad \varphi_g \leftarrow \varphi_g - r \nabla_{\phi_g} \mathcal{L}_{\text{RUL}}$
  \EndIf
    \State \textbf{Online variational inference (Birth and Merge):}
    \If {increment of the performance score $\Delta\mathcal{J} \ge 0 $}
        \State Perform birth move
        \State Perform merge move
        \State Re-infer $\{z_n\}$ and recompute $\mathcal{L}_{\text{RUL}}$ after accepted birth/merge moves
    \EndIf
\Until{Number of failure modes converge or maximum iterations reached}
\end{algorithmic}
\end{algorithm}

Note that the overall learning procedure is not purely derived from a posterior of a DPMM under a fixed prior and thus does not guarantee monotonic improvement of the ELBO. Instead, the inferred failure mode structure aims to balance statistical consistency in the sensor space (strict Bayesian optimality) with prognostic relevance for RUL prediction (practical usefulness). In the following sections, we compare the proposed method with a purely posterior-driven DPMM approach to highlight the impact of incorporating prognostic considerations into structural learning.

\section{Simulation studies}
\label{sec:simulation}

To validate the proposed methodology and demonstrate its robustness across different conditions, we conduct a series of numerical experiments using simulated system degradation signals. Section~\ref{sec:data_generation} outlines the procedure for generating the degradation data. Section~\ref{sec:stationary} evaluates the model’s performance under a four-mode scenario and investigates the impact of failure mode complexity and the sensitivity against hyperparameters in a stationary environment. Section~\ref{sec:nonstationary} shows the model's online learning capability under a non-stationary environment, i.e., where a new failure mode emerges.

\subsection{Data Generation}
\label{sec:data_generation}
\begin{figure}[!t]
\centering
\includegraphics[width=6in]{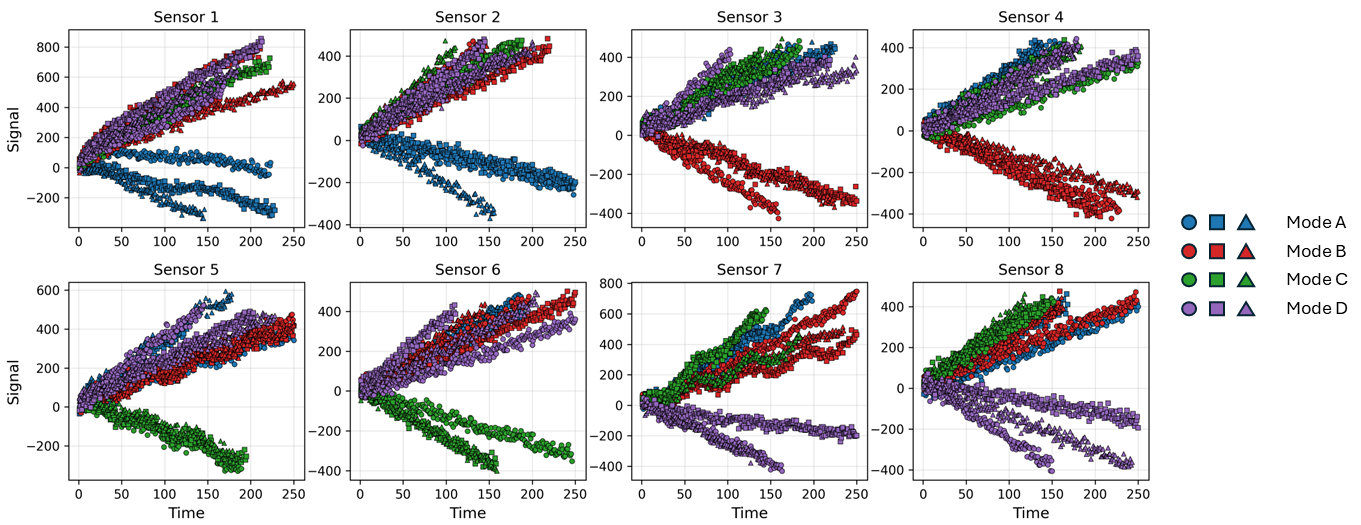}
\caption{Simulation sensor signal plots for the eight sensor signals of three systems (different shapes) per failure mode.}
\label{fig:simulation_dataset}
\end{figure}

We study a set of $N$ systems, each of which contains eight sensors ($S=8$) and is governed by one of four possible failure modes, $\mathcal{K}=\{A, B, C, D\}$. Figure~\ref{fig:simulation_dataset} shows the degradation of sensor signals of three randomly selected systems per failure mode, i.e., a total of 12 systems. 

Following the widely-used general degradation path model \citep{Lu1993g}, the underlying (noise-free) degradation trajectory of system $n$ under failure mode $z_n=j$ is modeled as a random-effect linear process:
\begin{equation}
\label{eq:degradation_model}
\begin{aligned}
    \eta_{n,j}(t) = \Gamma_{n,j,0} + \Gamma_{n,j,1}t, \quad j \in\{A, B, C, D\},
\end{aligned}
\end{equation}
where $\eta_{n,j}(t)$ corresponds to the true degradation status of system $n$ at time $t$ and \( \Gamma_{n, j} = (\Gamma_{n,j,0}, \Gamma_{n,j,1})^\top \) is a random effect vector drawn from a bivariate normal distribution that is specified for each failure mode $j$ \citep{kim2019generic}:
\begin{equation}
   \begin{aligned}
       \begin{pmatrix}
        \Gamma_{n, A,0} \\ \Gamma_{n, A,1}
        \end{pmatrix} \sim \mathcal{N}_2\left(
        \begin{pmatrix}
        -1.5 \\ 2.5
        \end{pmatrix},
        \begin{pmatrix}
        120 & 2 \\
        2 & 0.4
        \end{pmatrix}
        \right), \quad      
        \begin{pmatrix}
        \Gamma_{n, B,0} \\ \Gamma_{n, B,1}
        \end{pmatrix} \sim \mathcal{N}_2\left(
        \begin{pmatrix}
        -0.5 \\ 1.8
        \end{pmatrix},
        \begin{pmatrix}
        80 & 1 \\
        1 & 0.25
        \end{pmatrix}
        \right), \\
        \begin{pmatrix}
        \Gamma_{n, C,0} \\ \Gamma_{n, C,1}
        \end{pmatrix} \sim \mathcal{N}_2\left(
        \begin{pmatrix}
        -1.3 \\ 2.3
        \end{pmatrix},
        \begin{pmatrix}
        110 & 1.8 \\
        1.8 & 0.3
        \end{pmatrix}
        \right), \quad
        \begin{pmatrix}
        \Gamma_{n, D,0} \\ \Gamma_{n, D,1}
        \end{pmatrix} \sim \mathcal{N}_2\left(
        \begin{pmatrix}
        -0.8 \\ 2
        \end{pmatrix},
        \begin{pmatrix}
        90 & 1 \\
        1 & 0.4
        \end{pmatrix}
        \right).
   \end{aligned} 
\end{equation}
To ensure realistic, monotonically increasing degradation, realizations with $\Gamma_{n,j,1}\leq0$ are discarded and resampled. 

A failure threshold is set as \( L = 400 \), and the failure time for system $n$ is defined as:
\begin{equation}
    \tau_n = \inf\{t:\eta_{n,z_n}(t)\geq L\}=\frac{L - \Gamma_{n,z_n,0}}{\Gamma_{n,z_n,1}}.
\end{equation}
Sensor measurements are collected at integer time points $t=1,2,...,\lfloor\tau_n\rfloor$.
During training the prognostic module, for the $i$th sample (window) from system $n$ corresponding to measurements from time $t=i$ to $t=i+\xi-1$, its RUL label is defined as $y_{n,i}=\tau_n-(i+\xi-1)$.

Each sensor signal is constructed as a nonlinear function of time, the latent degradation status, system-specific random effects, and noise:
\begin{equation}
 l_{n,s,t} = \delta_s^{(1)} U_{n,s}^{(1)}t^{\delta_s^{(2)}}+\delta_s^{(3)}U_{n,s}^{(2)}f_s(t)+U_{n,s}^{(3)}+\delta_{z_n,s}^{(4)}\eta_{n, z_n}(t)+\varepsilon_{n,s}(t),
\end{equation}
where \( U_{n,s}^{(1)}, U_{n,s}^{(2)}, U_{n,s}^{(3)} \sim \text{Uniform}(0,30) \) introduce system-specific variability, \( \varepsilon_{n,s}(t) \sim \mathcal{N}(0,20^2) \) represents sensor noise, $f_s(t)$ is predefined temporal function (e.g., sine, cosine), $\delta_s^{(1)},\delta_s^{(2)},\delta_s^{(3)}$ control sensor-specific temporal behavior (Table \ref{tab:simulation_parameters}), and $\delta_{j,s}^{(4)}\in\{-1,1\}$ is a failure-mode-specific trend indicator to govern how degradation affects sensor $s$ under failure mode $j$ (Table \ref{tab:simulation_parameters_fm}).

\begin{table}[!t]
\centering
\begin{tabular}{@{} c c c c c p{4cm} @{}}
\hline
Sensor $s$ & $\delta_s^{(1)}$ & $\delta_s^{(2)}$ & $\delta_s^{(3)}$ & $f_s(t)$ \\
\hline
1 & 1     & 0.5  & 0.9 & sine(0.05t) \\
2 & 0.1   & 0.5  & 0.2  & 1 \\
3 & 2     & 0.01  & 1 & cosine(0.07t) \\
4 & 0.001 & 0.5  & 0  & --- \\
5 & 0.05  & 0.5    & 1 & sine(0.1t) \\
6 & 0.001 & 1.5  & 0.2  & sine(0.01t) \\
7 & 0.02  & 1.2  & 1.4 & cosine(0.1t) \\
8 & 0.01  & 0.5  & 0.14 & 1 \\
\hline
\end{tabular}
\caption{Sensor-specific parameters $\delta_s^{(1)},\delta_s^{(2)},\delta_s^{(3)}$ and temporal functions $f_s(t)$ used in the simulation data generation}
\label{tab:simulation_parameters}
\end{table}

\begin{table}[!t]
\centering
\begin{tabular}{@{} c c p{4cm} @{}}
\hline
Failure mode $j$ & $\delta_{j,s}^{(4)}$ \\
\hline
A & [-1, -1, +1, +1, +1, +1, +1, +1] \\
B & [+1, +1, -1, -1, +1, +1, +1, +1] \\
C & [+1, +1, +1, +1, -1, -1, +1, +1] \\
D & [+1, +1, +1, +1, +1, +1, -1, -1] \\
\hline
\end{tabular}
\caption{Failure-mode-specific sign parameters $\delta_{j,s}^{(4)}$ to control sensor trends under different failure modes}
\label{tab:simulation_parameters_fm}
\end{table}

This data generation enables the simulation of heterogeneous degradation paths and sensor patterns for comprehensive validation of the DPMM-RUL framework. 

\subsection{Stationary Environment}
\label{sec:stationary}

\subsubsection{Simulation Results}

We evaluate the performance of the proposed DPMM-RUL framework in a stationary environment with four failure modes. For each failure mode, 120 systems are included in the training set and 30 in the testing set. The number of failure modes is inferred using DPMM module while the RUL prediction module is trained jointly. Truncation and padding are used as a preprocessing strategy for the DPMM module.

\begin{figure*}[!t]
\centering
\includegraphics[width=4in]{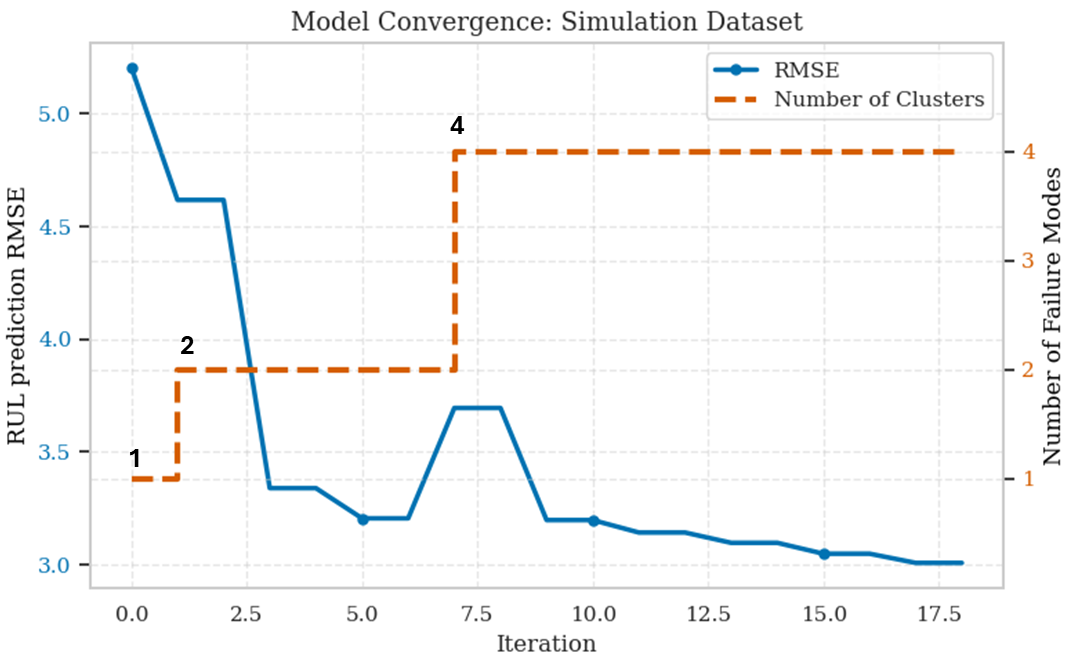}%
\caption{RMSE of RUL prediction and inference number of failure modes over iterations in Simulation dataset}
\label{fig:simulationstudy_result}
\end{figure*}

Figure~\ref{fig:simulationstudy_result} presents the evolution of the inferred number of failure modes and the RUL prediction RMSE over training iterations. We observe that the RMSE  for RUL prediction fluctuates when the inferred number of failure modes temporarily deviates from the true underlying number of failure modes. As learning progresses, the failure mode count stabilizes at four. At the same time, the stabilization of the failure mode identification leads to a consistent reduction in RUL prediction error. These results indicate that the mutual feedback between the DPMM and RUL modules allows the framework to converge to a coherent digital-twin representation of the system's degradation dynamics.

\subsubsection{Robustness Analysis}
\label{sec:robustness_analysis}

To systematically evaluate the robustness of the proposed method, we conduct experiments under varying (i) failure mode complexity, measured by the number of latent failure modes and (ii) similarity among degradation patterns. Specifically, we consider scenarios with two to four failure modes, and quantify their distinguishability using the average pairwise cosine similarity of the corresponding $\delta_j^{(4)}$-vectors listed in Table \ref{tab:simulation_parameters_fm}:

\begin{equation}
\label{eq:cosine_similarity}
    \mathrm{Similarity} \bigl( {\delta_j}^{(4)}, {\delta_{j'}}^{(4)} \bigr) 
= \frac{{\delta_j}^{(4)} \cdot {\delta_{j'}}^{(4)}}{\|{\delta_j}^{(4)}\|_2 \, \|{\delta_{j'}}^{(4)}\|_2}.
\end{equation}

Table~\ref{tab:rul_performance_sensiticity} reports, for each scenario, the number of detected failure modes, normalized mutual information (NMI) between inferred and true mode assignments, and the resulting RUL prediction RMSE. NMI quantifies the statistical dependence between predicted and ground truth failure modes, with values ranging from 0 (no mutual information) to 1 (perfect correlation).
Across all tested configurations, the proposed DPMM--RUL framework consistently identifies the correct number of failure modes and achieves perfect failure mode identification performance (NMI = 1.00), regardless of both the number of modes and their similarity level.

Importantly, increasing failure-mode similarity or complexity is generally expected to exacerbate both diagnostic ambiguity and prognostic uncertainty.
Nevertheless, the observed RMSE remains relatively stable across scenarios, indicating that accurate failure mode identification effectively mitigates the negative impact of mode overlap on RUL prediction.
These results demonstrate that the proposed framework maintains robust diagnostic and prognostic performance even under highly challenging degradation conditions.

\begin{table}[!t]
  \centering
  \begin{threeparttable}
  \caption{RUL prediction performance under different failure mode complexity and cosine similarities}
  \label{tab:rul_performance_sensiticity}
  \renewcommand{\arraystretch}{1.2}
  \begin{tabular}{@{}llccc@{}}
    \hline
    \textbf{Failure Modes} & \textbf{Similarity Level} & \textbf{Modes Detected} & \textbf{NMI} & \textbf{RMSE} \\
    \hline
    \multirow{2}{*}{2-Modes} 
      & Low ($-1.00$) & 2 modes & 1.00 & 12.94 \\
      & High ($0.75$)  & 2 modes & 1.00 & 16.87 \\
    \hline
    \multirow{2}{*}{3-Modes} 
      & Low ($0.17$) & 3 modes & 1.00 & 10.58 \\
      & High ($0.33$) & 3 modes & 1.00 & 11.07 \\
    \hline
    \multirow{2}{*}{4-Modes} 
      & Low ($0.17$) & 4 modes & 1.00 & 19.34 \\
      & High ($0.33$) & 4 modes & 1.00 & 21.53 \\
    \hline
  \end{tabular}
  \begin{tablenotes}
    \footnotesize
    \item Cosine similarity is calculated as the average pairwise cosine similarity between all $\delta$-vectors within each scenario.
  \end{tablenotes}
  \end{threeparttable}
\end{table}

To further understand the contribution of the proposed performance score $\mathcal{J}$, Table~\ref{tab:rul_performance_comparison} compares DPMM--RUL with three benchmark variants that differ in how the failure mode clustering structure is selected: (i) DPMM-joint fits a DP mixture model using a joint distribution $p(\mathbf{l}, y, z, \theta, \beta)$; (ii) DPMM--RUL (ELBO) selects the mixture structure solely by maximizing the ELBO; (iii) DPMM--RUL (RUL loss) fits the mixture model solely by minimizing the RUL prediction loss.

As shown in Table~\ref{tab:rul_performance_comparison}, the proposed DPMM-RUL based on the defined $\mathcal{J}$ score consistently recovers the correct number of failure modes, achieves perfect clustering performance (NMI = 1.00), and yields the lowest RMSE across all tested complexities. These results highlight that neither distributional fit nor prognostic accuracy alone is sufficient for reliable failure mode discovery. Rather, robust performance emerges only when diagnostic structure learning explicitly improves RUL prediction accuracy, which is regulated by cluster quality.

\begin{table}[!t]
  \centering
  \begin{threeparttable}
  \caption{RUL prediction performance comparison across different performance scores}
  \label{tab:rul_performance_comparison}
  \renewcommand{\arraystretch}{1.2}
  \begin{tabular}{@{}llccc@{}}
    \hline
    \textbf{Failure Modes} & \textbf{Method} & \textbf{Mode Detected} & \textbf{NMI} & \textbf{RMSE} \\
    \hline
    \multirow{4}{*}{2-Modes} 
      & DPMM-joint           & 1 Mode & 0.00 & 64.91 \\
      & DPMM-RUL (ELBO)      & 1 Mode & 0.00 & 15.50 \\
      & DPMM-RUL (RUL loss)  & 3 Modes & 0.82 & 14.92 \\
      & \textbf{DPMM-RUL ($\mathcal{J}$ score)} & 2 Modes & \textbf{1.00} & \textbf{12.94} \\
    \hline
    \multirow{4}{*}{3-Modes} 
      & DPMM-joint           & 1 Mode & 0.00 & 23.34 \\
      & DPMM-RUL (ELBO)      & 1 Mode & 0.00 & 15.69 \\
      & DPMM-RUL (RUL loss)  & 4 Modes & 0.90 & 14.25 \\
      & \textbf{DPMM-RUL ($\mathcal{J}$ score)} & 3 Modes & \textbf{1.00} & \textbf{11.07} \\
    \hline
    \multirow{4}{*}{4-Modes} 
      & DPMM-joint           & 1 Mode & 0.00 & 57.14 \\
      & DPMM-RUL (ELBO)      & 1 Mode & 0.00 & 18.56 \\
      & DPMM-RUL (RUL loss)  & 5 Modes & 0.94 & 19.56 \\
      & \textbf{DPMM-RUL ($\mathcal{J}$ score)} & 4 Modes & \textbf{1.00} & \textbf{19.34} \\
    \hline
  \end{tabular}
  \begin{tablenotes}
    \footnotesize
    \item The terms in parentheses correspond to the focus of performance scores.
  \end{tablenotes}
  \end{threeparttable}
\end{table}

\subsubsection{Sensitivity Analysis}
\label{sec:sensitivity_analysis}

We next investigate the sensitivity of the proposed framework to the trade-off coefficient $\omega$ in the $\mathcal{J}$ score. Table~\ref{tab:omega_sensitivity_analysis} summarizes the impact of $\omega$ under both low- and high-similarity settings for the three-mode scenario.
The sensitivity analysis demonstrates that the inferred number of failure modes and accuracy (NMI = 1.0) remain stable across varying $\omega$ and degradation similarity levels. In contrast, the RUL prediction error varies with respect to $\omega$. In practice, $\omega$ can be tuned using K-fold cross-validation based on RMSE of RUL prediction.

\begin{table}[!t]
  \centering
  \begin{threeparttable}
  \caption{Sensitivity analysis of the trade-off coefficient $\omega$ under varying failure mode complexities}
  \label{tab:omega_sensitivity_analysis}
  \renewcommand{\arraystretch}{1.2}
  \begin{tabular}{@{}lccccc@{}}
    \hline
    \textbf{Failure Modes} & $\mathbf{\omega}$ & \textbf{Similarity} & \textbf{Mode Detected} & \textbf{NMI} & \textbf{RMSE} \\
    \hline
    \multirow{4}{*}{3-Modes} 
      & $1\times10^{-4}$  & Low & 3 modes & 1.0 & 13.58 \\
      & $1\times10^{-3}$  & Low & 3 modes & 1.0 & 11.07 \\
      & $1\times10^{-2}$  & Low & 3 modes & 1.0 & 16.59 \\
      & $1\times10^{-1}$  & Low & 3 modes & 1.0  & 11.46  \\
    \hline
    \multirow{4}{*}{3-Modes} 
      & $1\times10^{-4}$  & High & 3 modes & 1.0 & 11.52 \\
      & $1\times10^{-3}$  & High & 3 modes & 1.0 & 10.58 \\
      & $1\times10^{-2}$  & High & 3 modes & 1.0 & 10.41 \\
      & $1\times10^{-1}$  & High & 3 modes & 1.0 & 12.59 \\
    \hline
  \end{tabular}
  \end{threeparttable}
\end{table}

\subsection{Online Learning in Non-stationary Environment}
\label{sec:nonstationary}
To evaluate the adaptability of the proposed framework under evolving operational conditions (i.e., introduction of a new failure mode), we consider a non-stationary environment in which the system initially operates under two failure modes, and an additional mode emerges at a later stage, simulating a real manufacturing scenario where new degradation behaviors appear due to process drift, component redesigns, or changes in operational load.

\begin{figure*}[!t]
\centering
\includegraphics[width=4in]{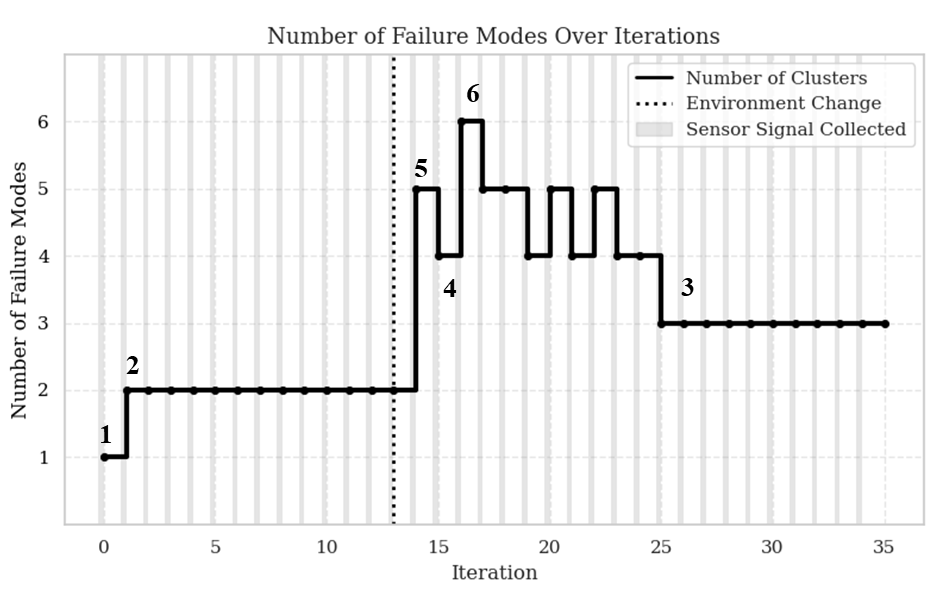}%
\caption{Number of inferred failure modes over iterations. The black dotted line shows when the new failure mode appears.}
\label{fig:simulation_result_nonstationary}
\end{figure*}

Figure~\ref{fig:simulation_result_nonstationary} shows the evolution of the number of inferred failure modes over iterations. The DPMM successfully detects the emergence of the third failure mode shortly after the change occurs, and performs a series of birth and merge moves, which results in three failure modes. This confirms the ability of the proposed framework to autonomously restructure its internal representation of multiple failure modes in response to previously unseen degradation patterns.

To show the predictive performance in a non-stationary environment, Table~\ref{tab:benchmark_comparison_2} compares the RUL RMSE of our method against several baselines that also study failure mode identification and system prognostics \citep{wang2024, li2023remaining, fu2025degradation}. To ensure reproducibility and a fair comparison, we include only methods for which we were able to obtain reproducible codes, either publicly released or provided directly by the original authors. These existing methods assume that there are two failure modes and cannot handle unseen failure modes. 

The table summarizes these methods based on whether they assume knowledge of the failure types of historical systems (labels) and the number of failure modes. \cite{wang2024} and \cite{li2023remaining} rely on both known failure-mode labels and a fixed number of failure modes, while \cite{fu2025degradation} assumes the number of modes is known but must infer the latent modes of historical systems. In contrast, DPMM-RUL does not assume either the types or the number of failure modes a priori. 

The results present that our DPMM-RUL framework based on the $\mathcal{J}$ score achieves the lowest RMSE and significantly outperforms existing methods. These results highlight the risks of prematurely assuming fixed and known failure-mode structures and demonstrate that the proposed framework is well-suited for non-stationary manufacturing environments.

\begin{table}[!t]
  \centering
  \begin{threeparttable}
  \caption{RUL prediction performance under increasing uncertainty in failure-mode information
  when the number of modes changes from 2 to 3}
  \label{tab:benchmark_comparison_2}
  \begin{tabular}{lcccc}
    \hline
    \multirow{2}{*}{\textbf{Method}} 
    & \multicolumn{2}{c}{\textbf{Failure mode known}} 
    & \multirow{2}{*}{\textbf{NMI}} 
    & \multirow{2}{*}{\textbf{RMSE (after shift)}} \\
    \cline{2-3}
     & \textbf{Label} & \textbf{Number} & & \\
    \hline
    \cite{li2023remaining} 
    & $\checkmark$ & $\checkmark$ & 0.00    & 60.24 \\
    \cite{wang2024}        
    & $\checkmark$ & $\checkmark$ & 0.73    & 63.93\\
    \cite{fu2025degradation} 
    & $\times$ & $\checkmark$ & --    & 18.02 \\
    DPMM-joint              
    & $\times$ & $\times$ & 0.00  & 19.82 \\
    DPMM-RUL (ELBO)         
    & $\times$ & $\times$ & 0.00  & 12.51 \\
    DPMM-RUL (RUL loss)     
    & $\times$ & $\times$ & 0.00  & 16.83 \\
    \textbf{DPMM-RUL ($\mathcal{J}$ score)} 
    & $\times$ & $\times$ & \textbf{1.00} & \textbf{12.41}  \\
    \hline
  \end{tabular}
  \begin{tablenotes}[flushleft]
    \footnotesize
    \item $\checkmark$ indicates that prior failure-mode information is assumed available,
    while $\times$ indicates that no such information is assumed. $-$ means the method does not identify failure modes.
    Rows are ordered from top to bottom by decreasing availability of prior information,
    corresponding to increasingly challenging settings.
  \end{tablenotes}
  \end{threeparttable}
\end{table}

\section{Case Study}
\label{sec:casestudy}

\subsection{Overview of Dataset}

The dataset used in this section is based on the Commercial Modular Aerospace Propulsion System Simulation (C-MAPSS) developed by NASA \citep{frederick2007user}, which simulates the degradation of turbofan engines under realistic operating conditions. It includes sensor signals collected from aircraft engines over their operational life, which can be used to monitor engine health and identify failure modes. For our case study, we focus on FD003, a sub-dataset featuring two distinct failure modes: high-pressure compressor (HPC) failure and engine fan (FAN) failure.

Each engine system in FD003 is equipped with 21 sensors and is observed across multiple operating cycles. The dataset is divided into a training set and a test set, each containing 100 systems. In the training set, each engine is monitored until failure occurs, and the actual failure time is recorded. In contrast, sensor monitoring data from systems in the test set are truncated before failure, with ground-truth RUL values provided as labels. Sensor descriptions are listed in Table~\ref{tab:cmapss}.

\begin{table}[h]
\caption{Detailed description of multiple sensors in CMAPSS dataset}
\centering
\begin{tabular}{c|l|c}
\hline
\textbf{Symbol} & \textbf{Description} & \textbf{Units} \\
\hline
T2 & Total temperature at fan inlet & $^\circ$R \\
T24 & Total temperature at LPC outlet & $^\circ$R \\
T30 & Total temperature at HPC outlet & $^\circ$R \\
T50 & T50 Total temperature at LPT outlet & $^\circ$R \\
P2 & Pressure at fan inlet & psia \\
P15 & Total pressure in bypass-duct & psia \\
P30 & Total pressure at HPC outlet & psia \\
Nf & Physical fan speed & rpm \\
Nc & Physical core speed & rpm \\
epr & Engine pressure ratio (P50/P2) & -- \\
Ps30 & Static pressure at HPC outlet & psia \\
phi & Ratio of fuel flow to Ps30 & pps/psi \\
NRf & Corrected fan speed & rpm \\
NRc & Corrected core speed & rpm \\
BPR & Bypass Ratio & -- \\
farB & Burner fuel-air ratio & -- \\
htBleed & Bleed Enthalpy & -- \\
Nf\_dmd & Demanded fan speed & rpm \\
PCNfR\_dmd & Demanded corrected fan speed & rpm \\
W31 & HPT coolant bleed & lbm/s \\
W32 & LPT coolant bleed & lbm/s \\
\hline
\end{tabular}
\label{tab:cmapss}
\end{table}

To retain the original signal structure, no additional feature preprocessing is applied. We follow the method introduced in \cite{chehade2018} to extract ground-truth failure mode labels, which are used to validate the accuracy of the failure mode identification module in our proposed algorithm.


\subsection{Results}
\label{sec:results_cmapss}

\begin{figure*}[!bt]
\centering
\includegraphics[width=3in]{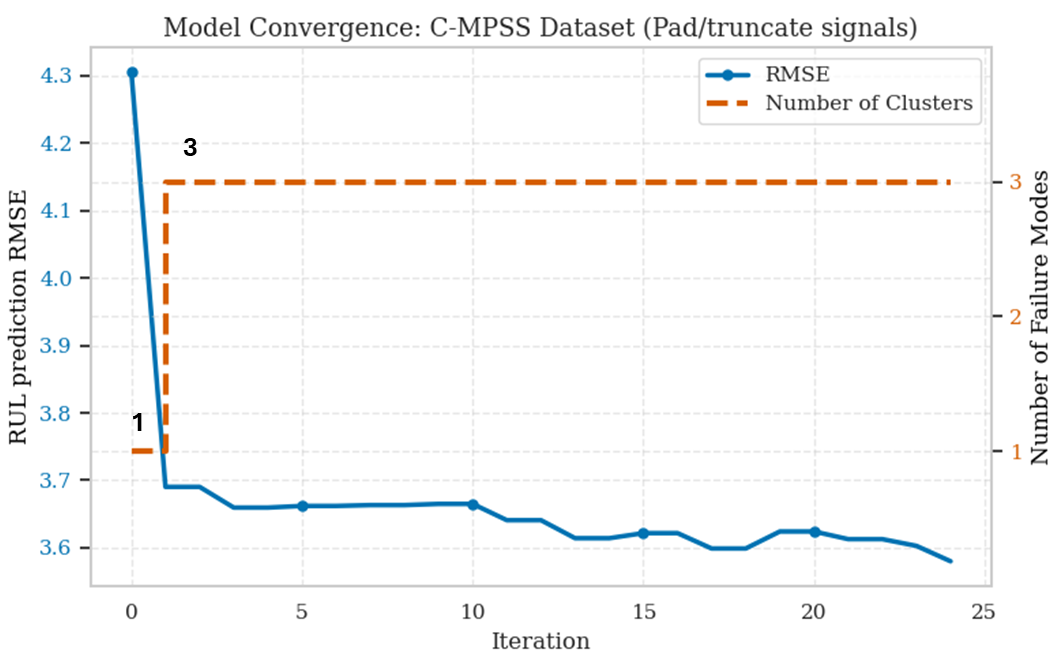}%
\hspace{1em}
\includegraphics[width=3in]{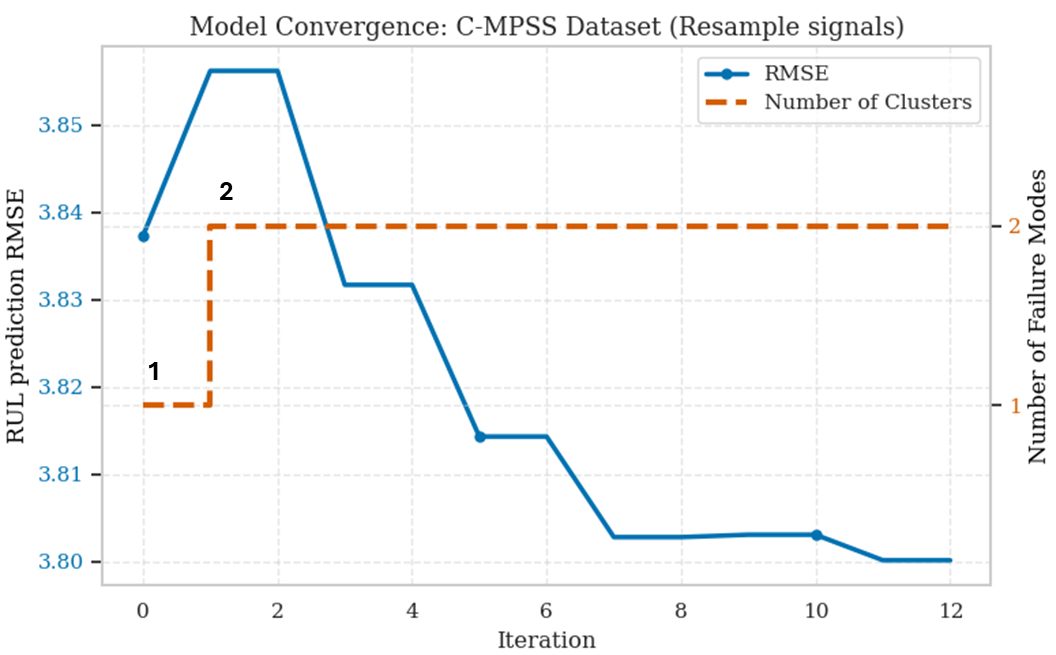}%
\caption{ RMSE of RUL prediction and inferred number of failure modes over iterations on FD0003 with (a) truncation and padding data preprocessing; and (b) temporal warping preprocessing.}
\label{fig:casestudy_result1}
\end{figure*}

Figure~\ref{fig:casestudy_result1} shows the inference and training dynamics of the proposed model on the FD003 sub-dataset of the C-MAPSS dataset. Similar to simulation studies, we observe that as training progresses, the RUL prediction accuracy steadily increases as the failure mode identification stabilizes.

\begin{figure*}[!t]
\centering
\includegraphics[width=4in]{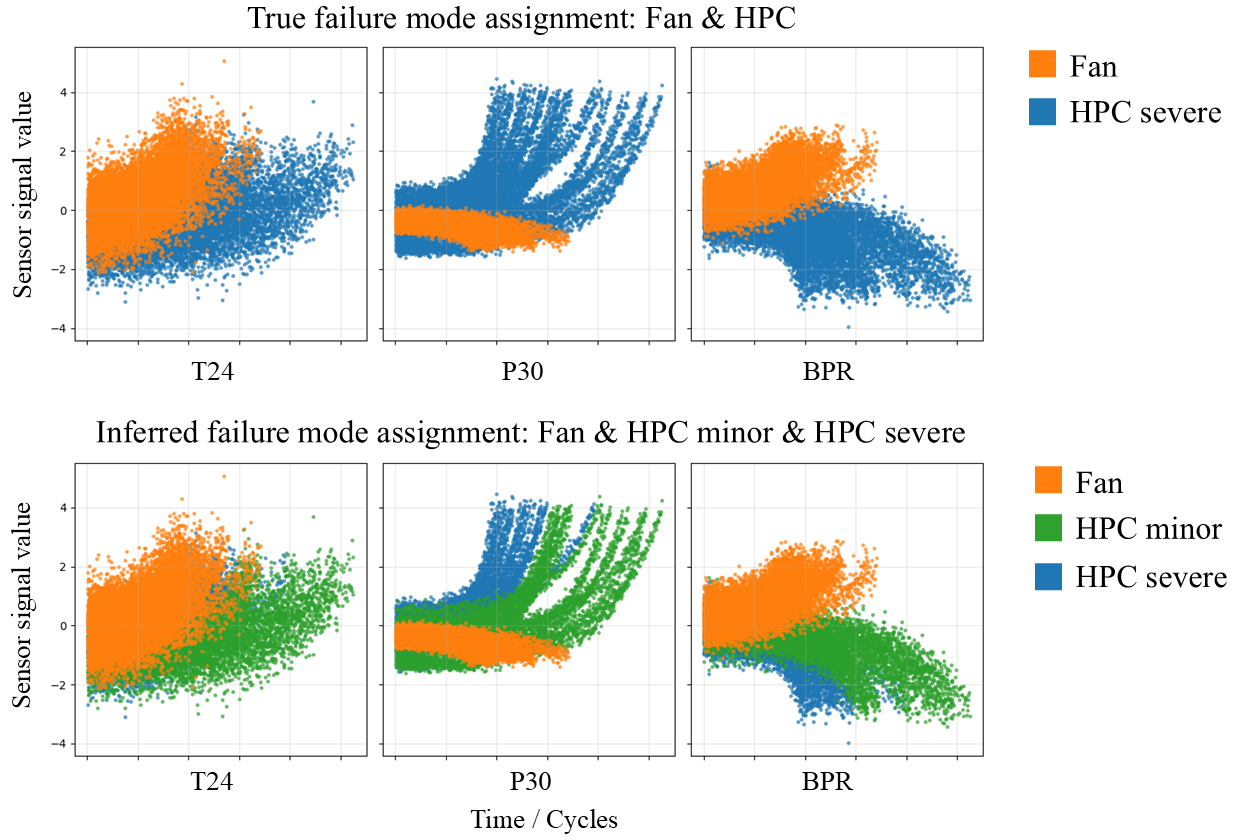}
\caption{Comparison between inferred and true failure modes in three representative sensors}
\label{fig:casestudy_result2}
\end{figure*}

Notably, when using truncation and padding as a preprocessing strategy (Figure~\ref{fig:casestudy_result1} (a)), the number of active failure modes inferred by the model stabilizes at three, even though the true system contains only two (i.e., HPC and FAN). A more detailed look at the inferred failure mode structure, as shown in Figure~\ref {fig:casestudy_result2}, reveals that the proposed methodology further subdivides the HPC failure mode into two subgroups, corresponding to minor and severe HPC degradation patterns. When these two HPC-related failure modes are combined, the results align precisely with the true failure mode structure. 

One possible explanation for this additional subdivision is the fixed-length preprocessing strategy described in Section \ref{sec:DPMM}. Under truncation and padding, trajectories from systems that experience earlier or more severe degradation contain longer padded segments. These segments may form an additional `statistical' cluster. To further examine this, we conduct the same analysis using a temporal warping preprocessing strategy. Specifically, for each training system, we define a normalized time index $\zeta=\frac{t}{T_n},\quad \zeta\in[0,1]$, and resample each multi-sensor trajectory using interpolation. Note that the predicted RUL is used for testing systems. This leads to the correct 2 failure modes as shown in Figure \ref{fig:casestudy_result1} (b).

This highlights that the selection of preprocessing should be guided by domain knowledge and the specific diagnostic objective of the user. In particular, it determines whether absolute time-to-failure should influence failure mode identification. If failure mechanisms are primarily distinguished by degradation trajectory shape, time warping may be preferable. Conversely, if degradation speed or lifetime duration is itself characterized by different failure behaviors, retaining original structure through truncation and padding may provide more meaningful information. The proposed DPMM–RUL framework accommodates either representation, while the semantic interpretation of inferred failure modes may differ accordingly.

Table~\ref{tab:benchmark_comparison} compares the RUL prediction accuracy of the proposed DPMM-RUL framework with the benchmark methods using C-MAPSS dataset. Despite imposing the least restrictive assumptions, our method significantly outperforms the benchmark that similarly lacks failure mode label information \citep{fu2025degradation} and performs competitively even against methods with access to full failure mode information \citep{wang2024, li2023remaining}. These results show that the DPMM-RUL can effectively recover the underlying mode structure and deliver accurate prognostics even in settings where failure mode information is entirely unavailable.

In terms of computational cost, the training time of the proposed DPMM-RUL ($\mathcal{J}$ score) remains comparable to other benchmark approaches. Importantly, the testing time of DPMM-RUL is negligible and comparable to other methods, indicating that the additional structural learning complexity does not affect deployment efficiency.

It is worth noting that the C-MAPSS dataset has a fixed and known number of failure modes, and thus does not fully stress the flexibility of our approach. As demonstrated in Section~\ref{sec:nonstationary}, the performance advantage of DPMM-RUL becomes more pronounced in scenarios where new or previously unseen failure modes emerge.

\begin{table}[!t]
  \centering
  \begin{threeparttable}
  \caption{RUL prediction performance comparison on C-MAPSS FD003
  under different availability of failure-mode information}
  \label{tab:benchmark_comparison}
  \begin{tabular}{lccccccc}
    \hline
    \multirow{2}{*}{\textbf{Method}} 
    & \multicolumn{2}{c}{\textbf{Failure mode known}} 
    & \multirow{2}{*}{\textbf{NMI}} 
    & \multirow{2}{*}{\textbf{RMSE}} 
    & \multirow{2}{*}{\shortstack{\textbf{Training}\\\textbf{time(s)}}}
    & \multirow{2}{*}{\shortstack{\textbf{Testing}\\\textbf{time(s)}}} \\
    \cline{2-3}
     & \textbf{Label} & \textbf{Number} &  &  &  &  &  \\
    \hline
    \cite{li2023remaining}        
    & $\checkmark$ & $\checkmark$ & 0.66 & \textbf{17.67} & 947.35 & 2.88 \\
    \cite{wang2024}               
    & $\checkmark$ & $\checkmark$ & 0.80 & 19.80 & 1891.49 & 0.06 \\
    \cite{fu2025degradation}      
    & $\times$ & $\checkmark$ & - & 24.19 & 6013.00 & 4.03 \\
    DPMM-joint      
    & $\times$ & $\times$ & 0.00 & 41.11 & 1335.41 & 0.73  \\
    DPMM-RUL (ELBO)      
    & $\times$ & $\times$ & 0.00 & 42.60 & 3789.42 & 0.06 \\
    DPMM-RUL (RUL Loss)  
    & $\times$ & $\times$ & 0.53 & 23.26 & 49703.73 & 0.07\\
    \textbf{DPMM-RUL ($\mathcal{J}$ score)}             
    & $\times$ & $\times$ & \textbf{0.82} & 18.56 & 5958.69 & 0.11 \\
    \hline
  \end{tabular}
  \begin{tablenotes}[flushleft]
    \footnotesize
    \item $\checkmark$ indicates that prior failure-mode information is assumed available,
    while $\times$ indicates that no such information is assumed. $-$ means the method does not predict the failure modes.
    Rows are ordered from top to bottom by decreasing availability of prior information,
    corresponding to increasingly challenging settings.
  \end{tablenotes}
  \end{threeparttable}
\end{table}

\section{Conclusion and Future Study}
\label{sec:conclusion}

This study presents a novel Bayesian nonparametric framework, DPMM-RUL, that jointly performs failure mode identification and RUL prognostics in multi-sensor systems without requiring prior knowledge of the type and number of failure modes.
By integrating a DPMM-based failure mode discovery module with a neural network-based RUL prediction module, the proposed approach overcomes critical limitations of existing methods that rely on fixed, predefined failure mode structures and fully labeled historical failures. 

Our framework enables adaptive identification of unknown and unseen failure modes while simultaneously predicting RULs. A key feature of DPMM-RUL is its iterative feedback mechanism. Failure mode identification informs RUL prediction, and the prognostic output in turn refines the inferred failure modes. Through this iterative feedback, the model adaptively infers and updates the number of failure modes as new systems are observed, enabling it to accommodate settings where the new failure mode may appear over time. 

Extensive numerical experiments on both simulated and the C-MAPSS datasets demonstrate the effectiveness of the proposed method in both accurate failure mode identification and RUL prognostics, particularly in scenarios involving unseen failure modes and incomplete prior knowledge. These findings suggest that the flexibility and adaptability of the proposed Bayesian nonparametric modeling offer substantial advantages for prognostics in complex manufacturing environments.

Future work will explore several other directions. First, to further reduce computational training time, the proposed method can be extended to apply amortized variational inference that replaces per-system variational parameter optimization with a learned inference network.
Second, adaptive model selection could be explored to determine the most effective model for different types of signal data. Another promising direction is to enhance the framework’s uncertainty quantification capability, providing more reliable confidence bounds for both failure mode identification and RUL prediction. This would further improve the interpretability and trustworthiness of the prognostic results in many applications.

\section*{Data Availability Statement}\label{data-availability-statement}

The data that supports the findings of this study is openly available in NASA Prognostics Center of Excellence Data Set Repository at https://data.nasa.gov/dataset/cmapss-jet-engine-simulated-data.

\if0\blind
{
\section*{Funding}
This work was supported by the National Science Foundation (NSF) grants 2530726 and 2450743.
} \fi

\if1\blind
{
} \fi

\bibliographystyle{apalike}
\spacingset{1}
\bibliography{ref}

\section*{Appendices}

\renewcommand{\thesubsection}{\Alph{subsection}}
\subsection{Evidence lower bound (ELBO) under the stick-breaking DPMM}
\label{sec:appendix}

Recall that our ELBO is defined as
\begin{align}
\mathcal{L}_{\text{ELBO}}(q)
&= \mathbb{E}_q[\log p(\mathbf{l}, z, \theta, \beta)]
- \mathbb{E}_q[\log q(z,\theta,\beta)] \notag \\
&= \mathbb{E}_q[\log F(\mathbf{l} \mid \theta,z)]
+ \mathbb{E}_q[\log p(z \mid \pi)]
+ \mathbb{E}_q[\log p(\pi \mid \alpha)]
+ \mathbb{E}_q[\log p(\theta \mid \lambda)] \notag \\
&\quad - \mathbb{E}_q[\log q(z)]
- \mathbb{E}_q[\log q(\beta)]
- \mathbb{E}_q[\log q(\theta)].
\end{align}
In the following, the explicit expectations under the NIW prior follow results in \cite{murphy2012machine}.

\begin{enumerate}

    \item $\mathbb{E}_q \left[ \log p(\mathbf{l} \mid \theta, z) \right]$:
    \begin{equation}
        \mathbb{E}_q \left[ \log p(\mathbf{l} \mid \theta, z) \right] = -\frac{1}{2}\sum_{n=1}^N \sum_{k=1}^K \hat{r}_{nk} \left(  \mathbb{E}_{q} \left[ \log |\boldsymbol{\Sigma}_{\theta_k}| \right] + \mathbb{E}_{q} \left[ (\mathbf{l}_n - \boldsymbol{\mu}_{\theta_k})^T \boldsymbol{\Sigma}_{\theta_k}^{-1} (\mathbf{l}_n - \boldsymbol{\mu}_{\theta_k}) \right] + \text{Constant} \right),
    \end{equation}
    \item $\mathbb{E}_q \left[ \log p(z \mid \pi) \right]$:
    \begin{equation}
        \mathbb{E}_q \left[ \log p(z \mid \pi) \right] = \sum_{n=1}^N\sum_{k=1}^K \hat{r}_{nk} \mathbb{E}_q \left[ \log \pi_k \right],
    \end{equation}
    where $\mathbb{E}[\log \pi_k] = \mathbb{E}[\log \beta_k]+\sum_{j=1}^{k-1} \mathbb{E}[\log (1-\beta_j)]$, $\mathbb{E}[\log \beta_k] = \gamma(\hat{\alpha}_{0, k}) - \gamma(\hat{\alpha}_{0,k} + \hat{\alpha}_{1,k})$, and $\mathbb{E}[\log (1-\beta_k)] = \gamma(\hat{\alpha}_{1, k}) - \gamma(\hat{\alpha}_{0,k} + \hat{\alpha}_{1,k})$. $\gamma$ is a digamma function.
    
    \item $\mathbb{E}_q \left[ \log p(\beta \mid \alpha) \right]$:
    \begin{equation}
        \mathbb{E}_q \left[ \log p(\beta \mid \alpha) \right] = \sum_{k=1}^{K-1} \left( (\alpha - 1) \log (1-\beta_k) + \log \alpha \right),
    \end{equation}
    where $\log \alpha = \log \Gamma (1+\alpha) - \log \Gamma (\alpha)$ and $\Gamma$ is a gamma function.
    \item $\mathbb{E}_q \left[ \log p(\theta | \lambda) \right]$:  \\
    \begin{equation}
        \begin{aligned}
        \mathbb{E}_q \left[ \log p(\theta \mid \lambda) \right] &= \sum_{k=1}^K \Bigg( \frac{\nu_0 + D + 2}{2} \mathbb{E}[\log |\boldsymbol{\Sigma}_k^{-1}|] - \frac{1}{2} \text{Tr}(\boldsymbol{\Psi}_0 \nu_k \boldsymbol{\Psi}_k^{-1}) \\
        &\quad - \frac{\kappa_0}{2} \left[ \frac{D}{\kappa_k} + \nu_k (\mathbf{m}_k - \mathbf{m}_0)^T \boldsymbol{\Psi}_k^{-1} (\mathbf{m}_k - \mathbf{m}_0) \right] + \text{Constant}\Bigg).
        \end{aligned}
    \end{equation}
    \item $\mathbb{E}_q \left[ \log q(\beta) \right]$:
    \begin{align}
        \mathbb{E}_q \left[ \log q(\beta) \right] = 
        & \sum_{k=1}^{K-1} [ (\hat{\alpha}_{0, k} - 1)(\gamma(\hat{\alpha}_{0, k}) - \gamma(\hat{\alpha}_{0,k} + \hat{\alpha}_{1,k})) + (\hat{\alpha}_{1, k} - 1)(\gamma(\hat{\alpha}_{1, k}) - \gamma(\hat{\alpha}_{0,k} + \hat{\alpha}_{1,k})) \\
        & + \log \Gamma(\hat{\alpha}_{0, k} + \hat{\alpha}_{1, k}) - \log\Gamma(\hat{\alpha}_{0, k}) - \log\Gamma(\hat{\alpha}_{1, k}) ].
    \end{align}
    \item 
    $\mathbb{E}_q \left[ \log q(z) \right]$:
    \begin{equation}
        \mathbb{E}_q \left[ \log q(z) \right] = \sum_{n=1}^N \sum_{k=1}^K \hat{r}_{nk} \log \hat{r}_{nk}.
    \end{equation}
    \item $\mathbb{E}_q \left[ \log q(\theta) \right]$:
    \begin{equation}
        \begin{aligned}
        \mathbb{E}_q \left[ \log q(\theta) \right] &= \sum_{k=1}^K \mathbb{E}_q[\log q(\mu_{\theta_k}\mid \Sigma_{\theta_k})] + \sum_{k=1}^K \mathbb{E}_q[\log q(\Sigma_{\theta_k})].
        \end{aligned}
    \end{equation}

\end{enumerate}


\subsection{Update equations for variational parameters}
    
Based on the explicit form of each expectation term derived in the variational inference framework, we present the update equations for the parameters $\hat{\alpha}_{0,k}$, $\hat{\alpha}_{1,k}$, $\hat{r}_{nk}$, and $\hat{\lambda}_k$.

\begin{enumerate}
\item Stick-breaking Parameters:
\begin{align}
\hat{\alpha}_{0,k} &= 1 + \sum_{n=1}^N \hat{r}_{nk}, \\[10pt]
\hat{\alpha}_{1,k} &= \alpha + \sum_{n=1}^N \sum_{j=k+1}^K \hat{r}_{nj}.
\end{align}

\item Variational Responsibilities:\\
\begin{equation}
\hat{r}_{nk} \propto \exp\left\{ \mathbb{E}_{q}[\log F(\mathbf{l}_n \mid \theta_k)] + \mathbb{E}_{q}[\log \pi_k] \right\},
\end{equation}

\item Failure Mode-Specific Parameters:\\
\begin{align}
    \kappa_k &=\kappa_0+N_k,\\
    \nu_k &= \nu_0+N_k, \\
    \mathbf{m}_k &= \frac{\kappa_0 m_0+N_k \bar{l}_k}{\kappa_0 + N_k}, \\ 
    \boldsymbol{\Psi}_k&=  \boldsymbol{\Psi}_0 + S_k + \frac{\kappa_0N_k}{\kappa_0 + N_k}(\bar{l}_k-m_0)(\bar{l}_k-m_0)^T,
\end{align}
where $N_k=\sum_{n=1}^N\hat{r}_{nk}$, $\bar{l}_k=\frac{1}{N_k}\sum_{n=1}^N\hat{r}_{nk} l_n$, and $S_k = \sum_{n=1}^N \hat{r}_{nk}(l_n-\bar{l}_k)(l_n-\bar{l}_k)^T$.
\end{enumerate}

\end{document}